\documentclass[useAMS, usenatbib, usegraphicx]{mn2e} 

\title[Galactic Winds and the CGM] 
{Galactic Winds in Cosmological Simulations of the Circumgalactic Medium}  

\author[P. Barai et al.] 
{Paramita Barai$^{1}$\thanks{E-mail: pbarai@oats.inaf.it}, 
Matteo Viel$^{1, 2}$, 
Stefano Borgani$^{3, 1, 2}$, 
Edoardo Tescari$^{4, 5}$, 
\newauthor 
Luca Tornatore$^{3}$, 
Klaus Dolag$^{6, 7}$, 
Madhura Killedar$^{3, 1}$, 
Pierluigi Monaco$^{3, 1}$, 
\newauthor 
Valentina D'Odorico$^{1}$, 
Stefano Cristiani$^{1, 2}$ 
\vspace{0.2cm} \\ 
$^{1}$ INAF - Osservatorio Astronomico di Trieste, Via G.B. Tiepolo 11, I-34143 Trieste, Italy \\ 
$^{2}$ INFN / National Institute for Nuclear Physics, Via Valerio 2, I-34127 Trieste, Italy \\ 
$^{3}$ Dipartimento di Fisica dell'Universit\`{a} di Trieste, Sezione di Astronomia, 
Via Tiepolo 11, I-34131 Trieste, Italy \\ 
$^{4}$ School of Physics, University of Melbourne, Parkville, VIC 3010, Australia \\ 
$^{5}$ ARC Centre of Excellence for All-Sky Astrophysics (CAASTRO) \\ 
$^{6}$ Universit\"{a}tssternwarte M\"{u}nchen, M\"{u}nchen, Germany \\ 
$^{7}$ Max-Planck-Institut f\"{u}r Astrophysik, Garching, Germany \\ 
}

\begin{document} 

\maketitle

\label{firstpage} 

\begin{abstract}

We explore new observationally-constrained sub-resolution models of galactic outflows  
and investigate their impact on the circumgalactic medium (CGM) 
in the redshift range $z = 2 - 4$. 
We perform cosmological hydrodynamic simulations, including star formation, 
chemical enrichment, and four cases of SNe-driven outflows: 
no wind (NW), 
an energy-driven constant velocity wind (CW), 
a radially varying wind (RVWa) where the outflow velocity 
has a positive correlation with galactocentric distance ($r$),     
and a RVW with additional dependence on halo mass (RVWb).     
Overall, we find that the outflows expel metal-enriched gas away from galaxies, 
significantly quench the star formation, 
reduce the central galactic metallicity and enrich the CGM. 
At $z = 2$, the radial profiles of gas properties around galaxy centers 
are most sensitive to the choice of the wind model 
for halo masses in the range $(10^{9} - 10^{11}) M_{\odot}$. 

We infer that outflows in the RVWb model are least effective, 
with results similar to the NW case, except that      
the CGM is enriched more. 
Moreover, we find that the models CW and RVWa are similar, 
both showing the impact of effective winds, with the following notable differences. 
RVWa causes a greater suppression of star formation rate at $z \leq 5$, 
and has a higher fraction of low-density ($\delta < 10$), 
warm-hot ($10^4 - 10^6$ K) gas than in CW. 
Outflows in CW produce a higher and earlier enrichment of some IGM phases than in RVWa. 
By visual inspection, we note that the RVWa model shows   
galactic disks more pronounced than all the other wind models. 
We predict that some observational diagnostics are more promising to 
distinguish between different outflow driving mechanisms in galaxies: 
$Z_C$ of the CGM gas at $r \sim (30 - 300) h^{-1}$ kpc comoving, and 
CIV fraction of the inner gas at $r < (4 - 5) h^{-1}$ kpc comoving. 

\end{abstract} 


\section{Introduction} 
\label{sec-intro} 

Energy feedback from star formation (SF) and powerful supernovae (SNe) 
explosions are believed to eject gas from galaxies driving outflows called 
{\it galactic winds}\footnote{Here, we use the terms {\it wind} and {\it outflow} 
synonymously, meaning continuous outward flow of gas from a galaxy, 
which might or might not escape depending on its velocity and galactic potential.} 
\citep[e.g.,][]{Johnson71, Mathews71, Larson74, Veilleux05}. 
SNe-driven outflows are detected in many observations at low 
redshifts \citep[e.g.,][]{Lynds63, McCarthy87, Bland88, Heckman90,
  Lehnert96, Strickland00, Strickland04, Rupke05}, reaching velocity
as high as $1000$ km/s \citep{Diamond-Stanic12}, and at high-$z$ 
\citep[e.g.,][]{Franx97, Pettini01, Adelberger03, Shapley03, Erb12, Newman12b}, 
extending up to distances of $60 - 130$ kpc physical \citep[e.g.,][]{Lundgren12}. 


We use the term circumgalactic medium (CGM) to denote the gas-phase structures, 
excluding star-forming gas, within the virial radius of galaxies. 
Galactic winds are considered to be the primary mechanism by which
metals are ejected out of star-forming regions in galaxies and
deposited into the CGM and the intergalactic medium (IGM) 
\citep[e.g.,][]{Larson75, Aguirre01, Scannapieco02, Aracil04, Fox07, 
Bouche07, Kawata07, Pieri07, Pinsonneault10, Gauthier12, Hummels12}. 
The IGM can also be metal-enriched by active galactic nuclei (AGN) feedback 
\citep[e.g.,][]{Khalatyan08, Germain09} and heated by blazars \citep[e.g.,][]{Puchwein12a}. 

SNe and starburst driven outflow is an important source of
feedback in galaxy evolution, and constitute a key ingredient of
current galaxy formation studies, both in 
hydrodynamic simulations (e.g., \citealt{SH03}, hereafter SH03, \citealt{Schaye10}) 
and in semi-analytical models \citep[e.g.,][]{Benson03, Bertone05}. 
The winds act by removing gas available to make stars, 
hence quench SF, and are argued to suppress
the formation of low-mass galaxies, flattening the low mass end of the 
luminosity function in simulations 
\citep[e.g.,][]{Theuns02, Rasera06, Stinson07}, which is closer to observations. 
Feedback is invoked to reproduce the realistic disk galaxies  
\citep[e.g.,][]{Weil98, Sommer-Larsen03, Governato04, Robertson04, Okamoto05}. 
These outflows are also argued to affect the large column density parts of 
the Lyman-$\alpha$ absorption line forest (LLSs, subDLAs and DLAs) 
seen in the spectra of distant quasars, which traces the IGM matter distribution 
\citep[e.g.,][]{McDonald05, Kollmeier06, Tescari11, Viel12}. 

The detailed internal physics underlying the origin and driving of
galactic outflows is complex, occuring on scales proper of the multiphase
structure of the interstellar medium (ISM) \citep[e.g.,][]{Heckman03,
  Stringer12}.  The gas is likely accelerated either by thermal
pressure \citep[e.g.,][]{Chevalier85}, radiation pressure
\citep[e.g.,][]{Murray05, Sharma11, Chattopadhyay12, Zhang12}, ram
pressure, cosmic rays \citep[e.g.,][]{Samui10, Uhlig12}; or a
combination of them \citep[e.g.,][]{Nath09, Everett10, Sharma12}. 
Cosmological simulations do not resolve the scales at which these
physical processes actually happen.  Outflows are hence implemented in
a phenomenological way in the simulations using sub-resolution
prescriptions and their interplay with the larger-scale environments
is investigated. 

Energy ejection by SNe and starburst in galaxies, often in the form of
powerful outflows, can be incorporated in a number of ways as a source
of feedback in the numerical scheme of cosmological simulations.  A
thermal feedback, where SNe energy is distributed as heating energy of
neighboring gas, is well-known to be ineffective 
\citep[e.g.,][]{Katz92, Friedli95, Steinmetz95, Katz96}, 
because the dense star-forming gas has high cooling rate, and the
injected thermal energy is radiated away quickly before it can
significantly impact the gas. 
Therefore depositing the SNe energy in the kinetic form 
is a more popular implementation in the literature, 
which has been shown to have significant feedback effects 
\citep[e.g.,][]{Navarro93, Mihos94, Cen00, Kawata01, SH03, DallaVecchia08, Dubois08, Oppe12}. 
In our current work we adopt this kinetic feedback by giving a velocity kick to the affected gas.

Alternative forms of SNe feedback have also been implemented for
example by considering that the affected gas undergoes adiabatic
evolution \citep[e.g.,][]{Mori97, Thacker00, Brook05}; by turning off
radiative cooling temporarily for part of the neighboring gas
\citep[e.g.,][]{Governato07, Piontek11}; or by distributing SNe energy to
hot and cold gas phases separately \citep[e.g.,][]{Marri03,
  Scannapieco06, Murante10}.  Note that few studies have proposed
prescriptions for efficient thermal feedback in SPH simulations
\citep[e.g.,][]{Kay03, DallaVecchia12}. 

Our kinetic feedback models (\S\ref{sec-num-WindFedbk}, \S\ref{sec-num-vwr}) 
are based on the so-called {\it energy-driven wind} scenario 
originally proposed by SH03, and subsequently used by others 
\citep[e.g.,][]{Tornatore04, Nagamine07, DallaVecchia08, Tescari09, Fabjan10, Barnes11}. 
Here, a fraction of SNe energy provides the outflow kinetic energy, 
and the wind speed is independent of galaxy mass or SF rate. 
If we consider the underlying physics, which occurs on scales 
orders of magnitude below the scales resolved in cosmological simulations, 
such an outflow is likely to be driven by the thermal pressure of SNe, 
and might be (more physically) called a {\it thermally-driven wind}. 

Other wind models in cosmological simulations generally involve 
a different formulation of outflow velocity and wind mass loading factor 
(described more in \S\ref{sec-num-WindFedbk}) 
in terms of galaxy properties (mass, velocity dispersion, SFR), 
sometimes suggested by observations. 
\citet{Oppenheimer06} implemented {\it momentum-driven wind}, 
\citep[also e.g.,][]{Tescari09, Tescari11}, 
using analytical prescriptions from \citet{Murray05} for 
momentum injection provided by radiation pressure of photons and SNe 
(or, {\it radiatively-driven wind}). 
\citet{Okamoto10} investigated models in which 
the outflow properties are dependent on the local velocity dispersion of the dark matter. 
\citet{Choi11} developed a multicomponent and variable velocity galactic outflow model. 
\citet{Puchwein12b} implemented a halo mass dependent energy-driven outflow model. 
\citet{Schaye10} compared a number of models 
of both energy-driven and momentum-driven outflows: 
few constant-mass loading and constant-velocity cases; 
mass loading and velocity dependent on gas density, 
gravitational potential, and halo circular velocity; 
wind particles temporarily decoupled hydrodynamically versus not decoupled; 
and thermal injection of SNe energy. 

In this paper we explore new wind models, where the 
galactic outflow speed is a function of radius (distance from galaxy 
center), as motivated by recent observations \citep{Steidel10}. 
We investigate both the cases of a mass-independent dependence on radius, 
and a halo mass dependent parametrization of the radially varying wind. 
To our knowledge, this is the first numerical implementation of both of these outflow models. 

This paper is organised as follows: 
we describe our numerical code and simulation setup in \S\ref{sec-numerical}, 
in \S\ref{sec-results} we present and discuss our results, 
while in \S\ref{sec-conclusion} we give a summary 
of the main findings and discuss possible future applications.

\section{Numerical Method} 
\label{sec-numerical} 

We use a modified version of the TreePM (particle mesh) - 
SPH (smoothed particle hydrodynamics) code {\sc GADGET-3}, 
whose first version was described in \citet{Springel01}. 
The public release of the code ({\sc GADGET-2}, \citealt{Springel05}) contains 
a time integration using energy and entropy conserving formulation of SPH \citep{Springel02}, 
uses fully adaptive smoothing lengths, 
and a standard SPH artificial viscosity prescription \citep{Monaghan97}. 
{\sc GADGET-3} includes a more efficient domain decomposition 
to improve the work-load balance over {\sc GADGET-2}. 

Some of the additional subgrid physics\footnote{By {\it subgrid} we mean 
{\it sub-resolution}, referring to physical processes occuring at length scales 
smaller than the resolved scales in our simulations.} 
included in the semi-public version of {\sc GADGET-3} code we use are 
outlined in \S\ref{sec-num-Subgrid}. The general galactic wind feedback model is
described in \S\ref{sec-num-WindFedbk}, our new outflow models in
\S\ref{sec-num-vwr}, and the code implementations in
\S\ref{sec-num-Implement}.  The simulations we perform are presented
in \S\ref{sec-num-sim}.

\subsection{Sub-resolution Physics: Radiative Processes, SF, Chemical Evolution} 
\label{sec-num-Subgrid} 

Radiative cooling and heating is based on the original implementation
of \citet{Katz96} and improved by adding metal-line cooling which is
implemented by adopting the cooling rates from the tables of
\citet{Wiersma09a}.  Net cooling rates are computed element-by-element
tracking 11 species: H, He, C, Ca, O, N, Ne, Mg, S, Si, Fe.  A
spatially-uniform time-dependent photoionizing background radiation is
considered from the cosmic microwave background (CMB) and the
\citet{Haardt01} model for the ultraviolet/X-ray background produced
by quasars and galaxies.  Contributions from the 11 elements are
interpolated as a function of density, temperature and redshift from
tables that have been pre-computed using the public photoionization
code CLOUDY \citep[last described by][]{Ferland98}, assuming the gas
to be dust free, optically thin and in (photo-)ionization equilibrium. 

SF and SNe feedback are implemented following the effective subresolution model by SH03. 
In this model, the physics of multiphase
structure of the ISM, on scales unresolved in cosmological
simulations, is modeled using spatially averaged properties describing
the medium on scales that are resolved. 
Gas particles with density above a limiting threshold, 
$n_{\rm SF} = 0.13$ cm$^{-3}$ (in units of number density of hydrogen atoms), 
are considered to contain two phases: cold condensed clouds that are in
pressure equilibrium with ambient hot gas. 
Each gas particle represents a region of the ISM, 
where the cold clouds supply the material available for SF. 
Star particles are collisionless, and are spawned from gas particles undergoing SF,
according to the stochastic scheme introduced by \citet{Katz96}.  We
allow a gas particle to spawn up to four generations of star
particles; therefore a typical star particle mass is about one-fourth
of the initial mass of gas particles. 

Stellar evolution and chemical enrichment feedback 
are implemented following the chemical evolution model of \citet{Tornatore07}. 
Production of 9 different metal species (C, Ca, O, N, Ne, Mg, S, Si, Fe) 
are accounted for using detailed yields from 
Type Ia SN (SN-Ia), Type II SN (SN-II), 
along with low and intermediate mass stars (LIMS) 
in the thermally pulsating asymptotic giant branch (TP-AGB) phase. 
Contributions from both SN-Ia and SN-II to thermal feedback are considered. 
Mass-dependent time delays with which different stellar populations release metals 
are included, adopting the lifetime function by \citet{Padovani93}. 
Different stellar yields are used: for SN-Ia taken from \citet{Thielemann03}, 
SN-II from \citet{Woosley95}, and LIMS from \citet{vandenHoek97}. 
The mass range for SN-II is considered to be $M / M_{\odot} > 8$, 
while that for SN-Ia originating from binary systems is $0.8 < M / M_{\odot} < 8$ 
with a binary fraction of $10\%$. 

We include a fixed stellar initial mass function (IMF) according to the formalism given by 
\citet{Chabrier03}, which is a power-law at $M / M_{\odot} > 1$ and has a log-normal form at masses below. 
However, we use power-law IMFs with different slopes    
over the whole mass range of $0.1$ to $100 M_{\odot}$, 
which mimics the log-normal form of \citet{Chabrier03} at lower masses, as tests indicate. 
In our model the functional form: 
$\phi \left( M \right) = K M^{-y}$, 
is composed of 3 slopes and normalizations: 
$y = 0.2$ and $K = 0.497$ for stellar masses $0.1 \leq M / M_{\odot} < 0.3$, 
$y = 0.8$ and $K = 0.241$ for $0.3 \leq M / M_{\odot} < 1$, and 
$y = 1.3$ and $K = 0.241$ for $1 \leq M / M_{\odot} < 100$. 
Stars within a mass interval $[8 - 40] M_{\odot}$ 
become SNe first before turning into black holes (BHs) at the end of their lives, 
while stars of mass $> 40 M_{\odot}$ are allowed to directly end in BHs. 

The chemical evolution model also incorporates mass loss through stellar winds and SNe explosions, 
which are self-consistently computed for a given IMF and lifetime function. 
A fraction of a star particle's mass is restored as diffuse gas 
during its evolution, and distributed to the surrounding gas particles. 
There is no AGN feedback in our simulations.

\subsection{Galactic Wind Feedback} 
\label{sec-num-WindFedbk} 

We use subgrid models for kinetic feedback via SNe-driven galactic
outflows, following the {\it energy-driven} prescription
(\S\ref{sec-intro}) (originally from SH03), where the mass and
energy carried away by outflows are regulated by two equations. 
The first equation relates the wind mass-loss rate $\dot{M}_w$ to the 
SF rate $\dot{M}_{\star}$:
\begin{equation} 
\label{eq-MdotWind} 
\dot{M}_w = \eta \dot{M}_{\star}. 
\end{equation} 
Here $\eta$ is the wind mass loading factor, or the wind efficiency. 
Observations have shown that in  
galaxies outflow mass-loss rates are comparable to or few times larger than SF rates 
\citep[e.g.,][]{Martin99, Pettini02, Bouche12, Newman12a}. 
Thus, following SH03, we adopt a constant $\eta = 2$ as the efficiency factor. 

The other equation relates the wind kinetic energy to a fixed fraction $\chi$ of SNe energy: 
\begin{equation} 
\label{eq-EnrgEquate} 
\frac{1}{2} \dot{M}_w v_w^2 = \chi \epsilon_{SN} \dot{M}_{\star}. 
\end{equation} 
Here $v_w$ is the wind velocity, 
$\epsilon_{SN}$ is the average energy released by SNe for each $M_{\odot}$ 
of stars formed under the instantaneous recycling approximation. 
Combining Eqs. (\ref{eq-MdotWind}) and (\ref{eq-EnrgEquate}), 
$v_w$ can be re-written as: 
\begin{equation} 
\label{eq-vW-EnrgDr} 
v_w = \left( \frac{2 \chi \epsilon_{SN}}{\eta} \right)^{1/2}. 
\end{equation} 
Unlike SH03, and following \cite{Tornatore07} and \cite{Tescari11},
we choose $v_w$ as a free parameter in our models, 
from which the effective $\chi$ can be computed using Eq. (\ref{eq-vW-EnrgDr}). 

For our adopted Chabrier power-law IMF (\S\ref{sec-num-Subgrid}), 
$\epsilon_{SN} = 1.1 \times 10^{49}$ erg $M_{\odot}^{-1}$. 
We initialize the wind energy fraction using $v_w = 400$ km/s, 
which is the constant velocity of our simulation run CW (\S\ref{sec-num-sim}). 
It corresponds to $\chi = 0.29$ of SNe energy being carried away by the wind. 

Outflow models implemented in the {\sc GADGET} code generally involve different 
scaling relations of $v_w$ and $\eta$ in terms of galaxy velocity dispersion, mass, and/or SFR. 
For the original energy-driven outflows, $v_w$ and $\eta$ are constant. 
The momentum-driven wind prescription by \citet{Oppenheimer06,Oppe08} has 
$v_w \propto \sigma \sqrt{(L/L_{\rm crit}) - 1}$ and $\eta \propto 1 / \sigma$, 
where $\sigma$ is the galaxy velocity dispersion, and 
$L / L_{\rm crit}$ is its luminosity in units of a critical value. 
In \citet{Choi11}'s model, 
$v_w \propto {\rm SFR}^{1/3}$ and $\eta$ is a function of galaxy stellar mass. 
\citet{Puchwein12b} assumed $v_w$ and $\eta$ to be proportional to halo mass. 

However, any dependence of $v_w$ on galactocentric radius (or distance from galaxy center) 
has not been explored before and this is what we are going to do in the present work.

\subsection{Wind Velocity Dependent on Galactocentric Distance, $v_w(r)$} 
\label{sec-num-vwr} 

The new subgrid wind model is motivated by the recent observational
studies of \citet{Steidel10}.  Studying the metal-enriched gas
kinematics in a region of $\sim 125$ kpc around star-forming
(Lyman-break) galaxies at redshifts $z = 2 - 3$, they are able to
reproduce their spectroscopic data using a simple model for outflows
and circumgalactic gas.  Considering a spherically-symmetric gas flow,
quantities are parameterized as a function of galactocentric distance,
or radius $r$.  The outflow acceleration is described as a power-law function 
of $r$:
\begin{equation} 
a(r) \propto r^{-\alpha} = \frac{dv(r)}{dt} = 
\left[ \frac{dv(r)}{dr} \right] \left( \frac{dr}{dt} \right) = v(r) \left[ \frac{dv(r)}{dr} \right]. 
\end{equation} 
Such a power-law acceleration have also been adopted recently by \citet{Gauthier12}. 
Integrating this equation with appropriate boundary conditions one can get 
the outward velocity $v(r)$, which is given in Eq. (16) of \citet{Steidel10}. 
We adopt a similar expression for the radially-varying velocity of galactic winds, namely: 
\begin{equation} 
\label{eq-vSteidel} 
v_w(r) = v_{\rm max} \left( \frac{r_{\rm min}^{1-\alpha} - r^{1-\alpha}} 
			     {r_{\rm min}^{1-\alpha} - R_{\rm eff}^{1-\alpha}} \right)^{0.5}. 
\end{equation} 
Here $r_{\rm min}$ is the distance from which the wind is launched and where the velocity is zero, 
$R_{\rm eff}$ represents the outer edge of gas distribution,  
$v_{\rm max}$ is the velocity at $R_{\rm eff}$ and $\alpha$ is a power-law index. 
In  Table 5 
of \citet{Steidel10}, the authors fit spectral line observations (of Ly$\alpha$, CIV, CII, SiII, SiIV) 
using different parameter values ranging between: 
$70 \leq R_{\rm eff} \leq 250$ kpc, and $650 \leq v_{\rm max} \leq 820$ km/s. 
Assuming $r_{\rm min} = 1$ kpc, 
they find that combinations of $\alpha$ within the range $1.15 \leq \alpha \leq 1.95$ 
together with another covariant parameter, can adequately reproduce observed ISM line profiles. 
In the absence of a unique set of parameter values, 
we choose the following for one of our simulations (run RVWa, \S\ref{sec-num-sim}): 
$r_{\rm min} = 1 h^{-1}$ kpc, $R_{\rm eff} = 100 h^{-1}$ kpc, $v_{\rm max} = 800$ km/s, and $\alpha = 1.15$. 

Such a trend of $v_w(r)$ correlating positively with $r$ has also been
found in other observations: for example \citet{Veilleux94}
parameterized outflow velocity as $v_w \propto r^n$ with $2 < n < 3$ for
the spiral galaxy NGC3079.

We note that a scenario in which the wind velocity is not unique 
but has instead any value (fast and/or slow outflow) could possibly result in 
faster ejecta at large distances from the galaxy center and slower ones 
closer to the center, since the former would have had more time 
to travel out to large distances. 
The model presented here could mimic such a behaviour. 

We explore a further halo mass dependent parametrization of our new wind model 
in run RVWb (\S\ref{sec-num-sim}). 
Observational studies of ultraluminous infrared galaxies 
at $z = 0.042 - 0.16$ by \citet{Martin05} 
detect a positive correlation of galactic outflow speed with galaxy mass. 
Their Figure 7 (bottom panel) shows that the outflow terminal velocities are always 
$2 - 3$ times larger than the galactic rotation speed. 
Inspired by these results, we assume that the velocity $v_{\rm max}$ is related to the 
circular velocity $v_{\rm circ}$ of the halo at the virial radius $R_{200}$ as: 
\begin{equation} 
\label{eq-vMax1} 
v_{\rm max} = 2 v_{\rm circ} = 2 \sqrt{ \frac{G M_{\rm halo}} {R_{200}} }. 
\end{equation} 
The halo mass $M_{\rm halo}$ and radius are related such that $R_{200}$ encloses 
a density $200$ times the mean density of the Universe at the simulation redshift $z$: 
\begin{equation} 
\label{eq-Mhalo} 
M_{\rm halo} = \frac{4 \pi}{3} R_{200}^3 
\left(200 \rho_{\rm crit} \Omega_{M,0}\right) \left(1 + z\right)^3. 
\end{equation} 
Here $\rho_{\rm crit} = 3 H_0^2 / (8 \pi G)$ is the present critical density. 
The assumption during code implementation is that the total group mass obtained by 
our group finder (\S\ref{sec-num-Implement}) is the virial halo mass 
and that it is collapsing at the simulation redshift. 
Using such a prescription the resulting dependence of maximum velocity on halo mass and redshift is:
\begin{equation} 
\label{eq-vMax2} 
v_{\rm max} \propto M_{\rm halo}^{1/3} \left(1 + z\right)^{1/2}. 
\end{equation} 
We also set the effective radius as: 
\begin{equation} 
R_{\rm eff} = R_{200} \propto M_{\rm halo}^{1/3} \left(1 + z\right)^{-1}, 
\end{equation} 
because $v_{\rm max}$ is the model wind velocity at $R_{\rm eff}$.

\subsection{Implementation in the {\sc GADGET-3} code} 
\label{sec-num-Implement} 

Our new wind model described in \S\ref{sec-num-vwr} requires an extra parameter: 
the distance of gas particles from their host galaxy center. 
We identify galaxies by running a Friends-of-Friends (FOF) group finder 
on-the-fly within our {\sc GADGET-3} simulations. 
Given a {\it linking length}, $L_{gr}$, the FOF algorithm finds a group by selecting 
all particles that lie within a distance $L_{gr}$ from any other particle in the group. 

Firstly, an average inter-particle distance, $L_{p}$, is calculated;
using the total mass, number of particles and the mean density.  The
linking length for the default on-the-fly FOF in {\sc GADGET-3} is
$L_{gr} / L_{p} = 0.16$, which is commonly used to find dark matter
(DM) halos.  However, sometimes the FOF can link together DM substructures
which actually belong to multiple galaxies with distinct stellar
structures, into a single halo.  Therefore, the stellar components of
the FOF groups are considered to be more stable and thereby identified as galaxies.

In order to select galaxies, we obtain stellar groups by modifying the
FOF group finder as follows:  we allow the FOF to first link over
stars as the primary particle type, and then link over (gas + DM)
particles as the secondary type.  To facilitate selection of stellar
groups, which are generally smaller in size, we use a linking length three
times smaller than the default one: $L_{gr} / L_{p} = 0.16 / 3$.  The FOF
is executed on-the-fly within a simulation at time intervals of a
multiplicative factor $1.001$ of the scale factor $a$, or, $a_{\rm
  next} / a_{\rm prev} = 1.001$.  All the groups above a minimum
length of $\geq 32$ particles are finally selected, which provide us
with a set of galaxies at a given simulation time. 
The code execution comprises of a FOF overhead of $(6 - 9) \%$ of the total processing time. 

Our preliminary tests indicate that the center-of-mass position of the
FOF groups are offset from the gas density peak location.  Hence we
decide not to consider the center-of-mass positions but to record the
position of the member gas particle with maximum SPH density in each
group:  this maximum gas density location is considered as the group
center.  We tag in the code the group (representing a galaxy) and the
group center each relevant gas particle belongs to. 
We define the galactocentric radius, $r$, as the distance of each gas particle from its group center. 

Assuming that outflows are generated from mass loss during SF, 
the implementation of launching the wind 
involves a probabilistic criteria similar to spawning 
star particles, as described in SH03.  Gas particles undergoing SF are
stochastically selected and kicked into wind, by imparting an one-time
$v_w$ boost. A probability is calculated in a timestep $\Delta t$ for
each multiphase gas particle:
\begin{equation} 
p = 1 - \exp \left( - \frac{\eta x \Delta t}{t_{\rm sfr}} \right). 
\end{equation} 
Here $x$ is the mass fraction of cold clouds and $t_{\rm sfr}$ is the
SF timescale, which depends on the parameters of the SF model.  A
random number is drawn in the interval [0, 1], and if it falls below
$p$ then the particle is given a wind kick.  If $\vec{v}$ is particle
velocity and $\phi$ its gravitational potential, then its velocity
after receiving wind kick is updated to:
\begin{equation} 
\label{eq-vNew} 
\vec{v}_{\rm new} = \vec{v}_{\rm old} + v_w \hat{y}. 
\end{equation} 
The direction of the unit vector $\hat{y}$ is set 
along $\left( \vec{v}_{\rm old} \times \vec{\nabla} \phi \right)$ 
or $- \left( \vec{v}_{\rm old} \times \vec{\nabla} \phi \right)$, 
randomly selected between the two. 
This makes the wind particles to be preferentially ejected along the 
rotation axis of the galaxy or perpendicular to the galaxy disk. 

In order to enable the outflow to escape from dense SF regions 
without affecting the SF, a new wind particle 
(gas particle just receiving a wind kick following Eq. \ref{eq-vNew})
is decoupled from hydrodynamic interactions, for a maximum duration 
$t_{\rm dec} = 0.025 t_{\rm H}(z)$, 
where $t_{\rm H}(z)$ is the Hubble time at a relevant simulation redshift. 
Then the wind particle does not enter hydrodynamic force
computation in the code, but is included in gravity and SPH density
calculation.  If the particle's density has fallen below 
$n_{\rm dec} = 0.25 n_{\rm SF}$, then the decoupling is stopped before
$t_{\rm dec}$, and full hydrodynamics are enabled again. 
In our simulations, the majority of wind particles 
end their decoupling phase following the density condition. 

Summarizing, the main free parameters in the wind model are $v_w$ and $\eta$, 
along with $t_{\rm dec}$ and $\rho_{\rm dec}$ for the decoupling prescription. 
The wind velocity is either constant (Eq. \ref{eq-vW-EnrgDr}, \S\ref{sec-num-WindFedbk}), 
or dependent on radius (Eq. \ref{eq-vSteidel}) and halo mass (Eq. \ref{eq-vMax2}). 

In this Section we have shown the actual implementation of different
forms of $v_w$, including radial and mass dependence of velocity which
are observationally motivated. We later study the implications such models 
have on the CGM and IGM properties in the simulated volume.

\subsection{Simulations} 
\label{sec-num-sim} 

The series of simulations we perform are listed in
Table~\ref{Table-Sims}.  A cosmological volume, with periodic boundary
conditions, is evolved starting with an equal number of DM and gas 
particles from $z = 99$, where the initial conditions have been generated 
using the {\sc CAMB}\footnote{http://camb.info/} software \citep{lewisetal}, 
up to $z = 2$. 
A concordance flat $\Lambda$CDM model is used with the following parameters: 
$\Omega_{M, 0} = 0.2711, \Omega_{\Lambda, 0} = 0.7289, 
\Omega_{B, 0} = 0.0463, n_{\rm S} = 0.96, \sigma_8 = 0.809, 
H_{0} = 70.3$ km s$^{-1}$ Mpc$^{-1}$, in agreement with 
recent observations of the cosmic microwave background radiation, 
weak gravitational lensing, Lyman-$\alpha$ forest and 
mass function evolution of galaxy clusters 
\citep[e.g.,][]{Lesgourgues07, Vikhlinin09, Komatsu11}. 

%



\begin{table*} 
\begin{minipage}{10cm} 
\caption{ 
Simulation Parameters. 
Column 1: Name of simulation run. 
Names ending with "t" are smaller boxsize runs SB, 
and LB are runs with larger boxsize, all of which are stopped at $z \sim 2$. 
Column 2: $L_{\rm box}$ = Comoving side of cubic simulation volume. 
Column 3: Specifications of galactic wind feedback model. 
Parameters of SB runs: 
Number of gas and DM particles in the initial condition, 
$N_{\rm part} = 2 \times 128^3$. 
Mass of gas particle (which has not undergone any star-formation), 
$m_{\rm gas} = 7.66 \times 10^{5} h^{-1} M_{\odot}$. 
Gravitational softening length (of all particle types), 
$L_{\rm soft} = 0.98 h^{-1}$ kpc. 
In runs RVWat and RVWa, parameters of radially varying wind model (\S\ref{sec-num-vwr}): 
$r_{\rm min} = 1 h^{-1}$ kpc, $R_{\rm eff} = 100 h^{-1}$ kpc, $v_{\rm max} = 800$ km/s, $\alpha = 1.15$. 
In runs RVWbt and RVWb, 
parameters of radially varying wind model dependent on halo mass (\S\ref{sec-num-vwr}): 
$v_{\rm max} = 2 \sqrt{ G M_{\rm halo} / R_{200} }$, and $R_{\rm eff} = R_{200}$. 
Parameters of LB runs: 
$N_{\rm part} = 2 \times 320^3$, 
$m_{\rm gas} = 6.13 \times 10^{6} h^{-1} M_{\odot}$, 
$L_{\rm soft} = 1.95 h^{-1}$ kpc. 
} 


\label{Table-Sims} 
\begin{tabular}{@{}ccc}

\hline

Run  & $L_{\rm box}$ & Galactic Wind Feedback \\ 
Name & [$h^{-1}$ Mpc] \\ 

\hline

\multicolumn{3}{c}{Smaller-Box Runs : SB} \\ \\ 

NWt   & $5$ & No Wind \\  

CWt   & $5$ & Energy-driven constant-velocity $v = 400$ km/s \\ 

RVWat & $5$ & Radially varying with fixed parameters \\ 

RVWbt & $5$ & RVW with halo mass dependent parameters \\ \\

\multicolumn{3}{c}{Larger-Box Runs : LB} \\ \\ 

NW   & $25$ & No Wind \\   

CW   & $25$ & Energy-driven constant-velocity $v = 400$ km/s \\   


RVWa & $25$ & Radially varying with fixed parameters \\   

RVWb & $25$ & RVW with halo mass dependent parameters \\ 


\hline
\end{tabular} 

\end{minipage}
\end{table*}



The smaller box-size $L_{\rm box} = 5 h^{-1}$ Mpc comoving series (or, box {\it SB}) has
$N_{\rm part} = 2 \times 128^3$ DM and gas particles in the initial
condition.  This is the higher resolution series with gas particle
mass $m_{\rm gas} = 7.66 \times 10^5 h^{-1} M_{\odot}$, and DM
particle mass $m_{DM} = 4.49 \times 10^6 h^{-1} M_{\odot}$. 
The Plummer-equivalent softening length for gravitational forces is set to 
$L_{\rm soft} = 0.98 h^{-1}$ kpc comoving. 
In the larger box-size $L_{\rm box} = 25 h^{-1}$ Mpc comoving
series (or, box {\it LB}) the numbers are: $N_{\rm part} = 320^3, m_{\rm gas} = 6.13
\times 10^6 h^{-1} M_{\odot}, m_{DM} = 3.59 \times 10^7 h^{-1}
M_{\odot}, L_{\rm soft} = 1.95 h^{-1}$ kpc. 
The minimum gas smoothing length is set to a fraction $0.001$ of $L_{\rm soft}$ 
in all the simulations. 
However the minimum smoothing which is 
actually achieved in the simulations depend on the resolution, 
and in our runs the gas smoothing lengths went down to $\sim 0.2 L_{\rm soft}$. 
The second set of simulations, at slightly worse resolution compared to the first set, 
has been chosen in order to increase the statistics, 
since more number of halos and more massive halos are expected to form in the larger box. 

In each series four runs are performed incorporating the same non-wind
subgrid physics described in \S\ref{sec-num-WindFedbk}, and
investigating different galactic wind models: \\ (1) NW: no wind;
\\ (2) CW: energy-driven wind with constant velocity
(\S\ref{sec-num-WindFedbk}, Eq. \ref{eq-vW-EnrgDr}) $v_w = 400$ km/s;
\\ (3)  RVWa: radially varying wind with fixed parameters
(\S\ref{sec-num-vwr}) $r_{min} = 1 h^{-1}$ kpc, $R_{eff} = 100 h^{-1}$
kpc, $v_{max} = 800$ km/s, $\alpha = 1.15$; and \\ (4) RVWb: radially
varying wind with parameters dependent on halo mass. \\ 
The smaller box-size runs have an extra {\it t} appended to the names. 
Few results are presented for both the box sizes, namely SB and LB runs. 
While the remaining results are shown only for one box-size, 
with the reason given in the relevant parts of \S\ref{sec-results} next. 


\begin{figure*}  
\centering 
\includegraphics[width = 0.8 \linewidth]{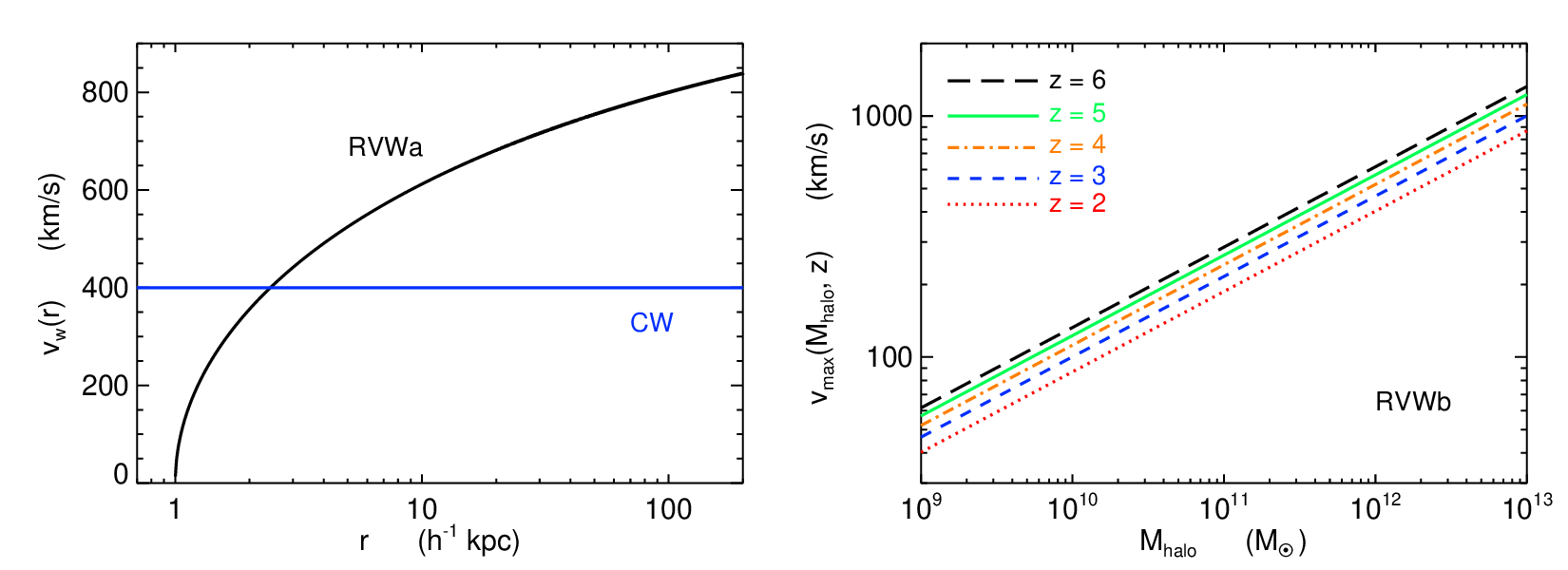} 
\caption{ 
{\it Left panel:} Black curve is the radial profile of outflow velocity 
as formulated in Eq.~(\ref{eq-vSteidel}),     
which is implemented in our hydrodynamic simulations motivated by observations of \citet{Steidel10}. 
The parameter values are for run RVWa (Table \ref{Table-Sims}): 
$r_{\rm min} = 1 h^{-1}$ kpc, $R_{\rm eff} = 100 h^{-1}$ kpc, 
$v_{\rm max} = 800$ km/s, and $\alpha = 1.15$. 
The blue horizontal straight line denotes the constant value of velocity, $v = 400$ km/s, in run CW. 
{\it Right panel:} Velocity at $R_{\rm eff}$ as a function of halo mass, 
shown in Eqs.~(\ref{eq-vMax1}) and (\ref{eq-vMax2}), for different redshifts of collapse. 
Such a formulation is implemented in run RVWb motivated by observational studies of \citet{Martin05}. 
} 
\label{fig-Vinput_vs_r_vs_Mhalo} 
\end{figure*} 



The new velocity prescriptions are plotted in Figure~\ref{fig-Vinput_vs_r_vs_Mhalo}. 
The black curve in the left panel shows the outflow velocity 
as a function of galactocentric radius for the parameters in run RVWa. 
For comparison, 
the blue horizontal line denotes the constant value of velocity, $v = 400$ km/s, used in run CW. 
The right panel depicts the parameterization of $v_{\rm max}$, or the 
velocity at $R_{\rm eff}$ in run RVWb, as a function of halo mass for different redshifts of collapse. 

We have analyzed the carbon content of the gas in the box, 
since carbon is one of the most abundant heavy element in the Universe, 
and the spectral lines produced by ionized carbon are relatively easy to observe. 
The metallicity of carbon, $Z_C$, 
is computed as the ratio of carbon mass to the total particle mass for each gas particle. 
Abundance ratios are expressed in terms of the Solar metallicity, 
which is $Z_{C, \odot} = 0.002177$ (mass fraction of carbon in Sun) 
derived from the compilation by \citet{Asplund05}.

\section{Results and Discussion}
\label{sec-results} 

\subsection{Outflow Velocity} 
\label{sec-res-OutVel} 

\begin{figure*} 
\centering 
\includegraphics[width = 0.9 \linewidth]{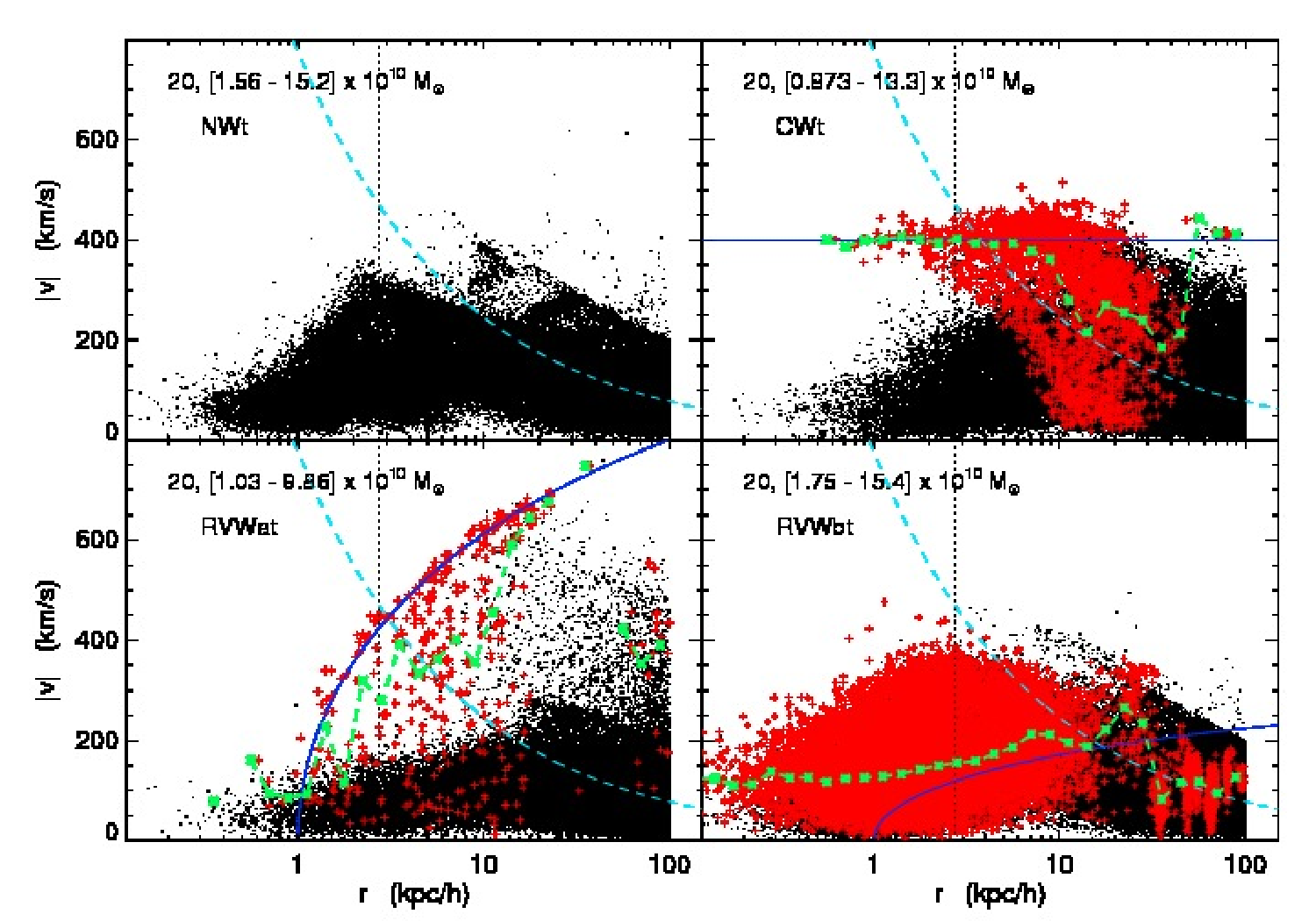} 
\caption{ 
Velocity magnitude of the gas as a function of distance from galaxy center 
at $z = 2.44$ in the 
SB runs with different wind models: 
no wind (top-left), energy-driven constant-velocity wind (top-right), 
radially varying wind with fixed parameters (bottom-left), and 
radially varying wind with parameters dependent on halo mass (bottom-right). 
All the gas particles within $100 h^{-1}$ kpc from the group center 
are shown as black points. 
In the runs with wind, the red plus symbols are the wind-phase particles 
(described in \S\ref{sec-res-OutVel}), whose median velocity in radial bins 
is denoted by the green asterisks joined by the green dashed curve. 
The blue solid curve represents the subgrid wind velocity implemented in the code 
in each case; in the last panel (run RVWbt) the blue solid line refers to 
a representative halo mass $M_{\rm halo} = 10^{11} M_{\odot}$. 
Each panel shows the stacked result of the 20 most-massive galaxies 
having total mass within the range specified in the panels. 
The vertical black dotted line denotes the resolved length scale of 
Newtonian gravitational forces ($2.744 h^{-1}$ kpc in the SB runs), 
which is $2.8$ times $L_{\rm soft} = 0.98 h^{-1}$ kpc comoving. 
The cyan dashed curve shows the escape velocity, 
$v_{\rm esc} = \sqrt{2 G M_{\rm halo} / r}$, for the representative halo mass. 
}
\label{fig-VelGas_vs_Dist} 
\end{figure*}

The different implementations of the outflow velocity in the  
SB runs are shown in the panels of 
Fig.~\ref{fig-VelGas_vs_Dist}: NWt (top-left), CWt (top-right), RVWat
(bottom-left), and RVWbt (bottom-right). 
Here the smaller boxsize is chosen for visualization purposes, 
to unambiguously distinguish the wind velocity from the remaining gas 
in the most-massive galaxy within the box volume which has the largest SFR. 
The trends for the LB runs are the same as the SB results shown here; 
however for galaxies of this mass range there are fewer wind particles because of lower SFR, 
since the LB runs produce more massive galaxies with higher SFR 
in a larger cosmological volume than SB. 
The velocity magnitude of the gas particles is plotted in Fig.~\ref{fig-VelGas_vs_Dist} 
as a function of distance from center of galaxy at $z = 2.44$. 
As described in \S\ref{sec-num-Implement}, 
galaxies are identified on-the-fly in the simulations using a FOF
algorithm linking over star particles as the primary type using a linking length 
3 times smaller than the default one, 
and the galaxy center corresponds to the location of the maximum gas density. 


\begin{figure*} 
\centering 
\includegraphics[width = 0.9 \linewidth]{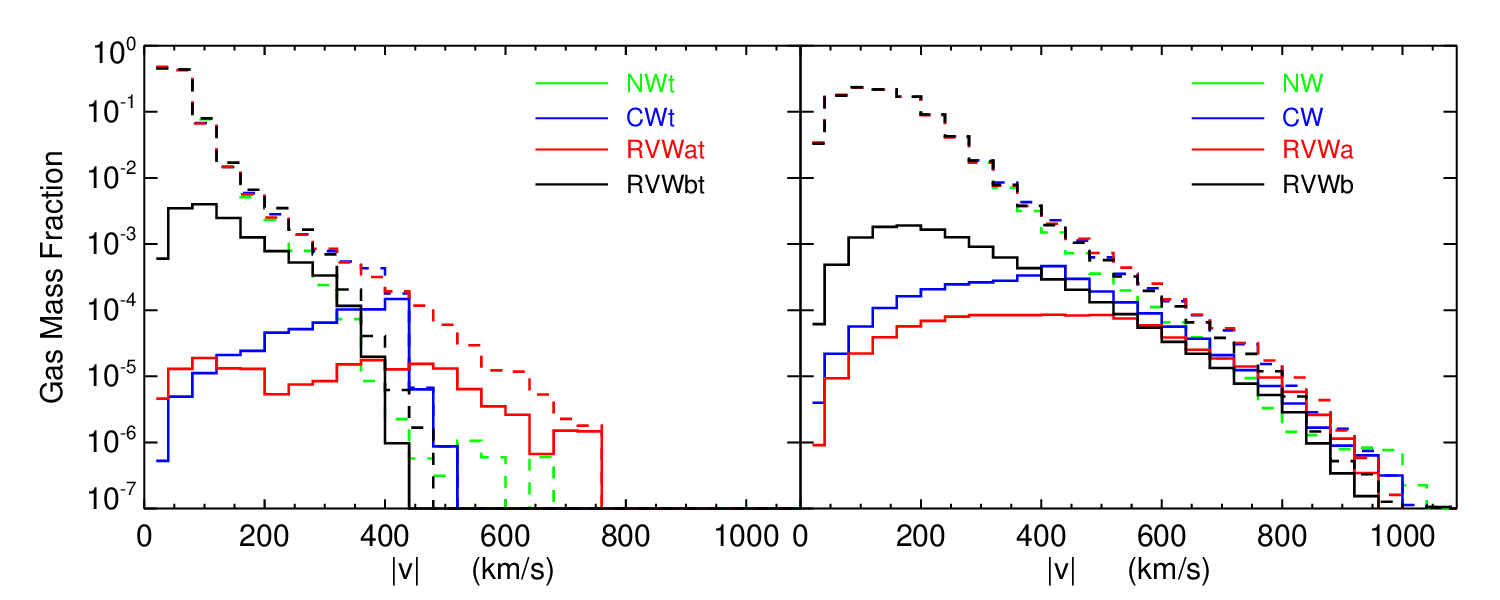} 
\caption{ 
Velocity magnitude histogram of wind particles (solid), and all gas particles (dashed) 
at $z = 2.91$, showing the mass fraction of gas per velocity bins.    
The SB runs are in the left panel and LB runs are in the right panel, 
in each case the histogram is normalized to the total gas mass in the simulation at the redshift considered. 
} 
\label{fig-VelHist} 
\end{figure*} 



All the gas particles lying within a radius of $100
h^{-1}$ kpc from group center are shown as black points in
Fig.~\ref{fig-VelGas_vs_Dist}.  The particles which have recently
received a wind kick velocity can be tracked, for the duration of
their hydrodynamic decoupling, through their positive decoupling time
(\S\ref{sec-num-Implement}).  We call these the {\it wind-phase
  particles} and we mark them with the red plus symbols.  The median 
wind-phase particle's velocity in radial bins is plotted as the green
asterisks joined by the green dashed curve.  The input subgrid wind
velocity ($v_w$) implemented in the code is instead represented by the
blue solid curve.  In order to mark the differences clearly, 
in each run results are stacked over the 20 most-massive galaxies 
having total halo (DM + gas + stars) mass between 
$M_{\rm halo} = (0.9 - 15) \times 10^{10} M_{\odot}$, the exact 
range for each case is written in the respective panel. 
The escape velocity, $v_{\rm esc} = \sqrt{2 G M_{\rm halo} / r}$, 
for a representative halo mass $M_{\rm halo} = 10^{11} M_{\odot}$, 
is shown by the cyan dashed curve. 
At a given radius, particles moving faster than $v_{\rm esc}$ are unbound to the halo. 

Run NWt gives the radial structure of gas velocity without any wind, 
reaching a maximum of $v \sim 400 - 500$ km/s around galaxies. 
The other 3 runs have winds implemented and present a distribution of
red points.  Wind particles are kicked over a radial range from the
center up to $r \sim (30 - 50) h^{-1}$ kpc, marking the size of
star-forming galaxy disk.  These particles gain an additional velocity
of $v_w$ during their wind kick (\S\ref{sec-num-Implement}).  Runs CWt
and RVWat demonstrate this as some red points clustered around or
somewhat above the blue curve.  Afterward they move away from galaxy
center and get decelerated, causing a fraction of the red points to
fall well below $v_w$.  Therefore, when considering the wind
particles at given $r$, the median velocity (green dashed curve) is
somewhat lower than $v_w$ (blue).  Run CWt has particles kicked by
$v_w = 400$ km/s at all $r$, while in our new implementation RVWat the
upper envelope of the red points follow the $v_w(r)$ radial profile of
Eq.~\ref{eq-vSteidel}.  In run RVWbt (bottom-right panel), the blue
solid curve shows $v_w$ for the halo mass $M_{\rm halo} = 10^{11} M_{\odot}$. 
This turns out to be a
weak wind, because most of the gas is moving at velocity above the
blue curve at $r \leq 10 h^{-1}$ kpc.  Hence the wind particles do not
move outward appreciably, and end up following a distribution very
close to the remaining gas (black points). 

Fig.~\ref{fig-VelGas_vs_Dist} also demonstrates that even though wind 
particles are kicked only one time according to the original
prescription of SH03, since this procedure is done at every timestep
(stochastically selecting from all the star-forming particles), the
desired $v_w$ is reproduced by multiple gas particles at any given
time.  The wind particles slow down, but in the next timesteps more particles
are kicked out with $v_w$, as long as there is SF, thus roughly
mimicking a continuous outflow of gas moving with given subgrid input
speed around a galaxy. 

Velocity magnitude histograms of gas particles at $z = 2.91$ are
plotted in Fig.~\ref{fig-VelHist}, showing mass fraction in velocity
bins; wind (gas) particles are represented by solid (dashed) curves.
In each case the histogram is normalized to the total gas mass in the 
respective simulation. 
The SB runs are shown in the left panel 
where the maximum particle velocity is $\sim 700$ km/s, and 
LB runs in the right panel where particles reach $\sim
1100$ km/s because of higher-mass halos forming in a larger box.

Runs CWt and CW have peaks in their wind particle velocity histogram at $v \sim 400$ km/s, 
since in this case $v_w$ is kept constant at this value.  The wind
velocity histogram is much flatter in runs RVWat and RVWa with no
well-defined peak, because $v_w$ has a range of values depending on
galactocentric distance.  Runs RVWbt and RVWb also have a peak wind
velocity, which is at a lower value $v \sim 100 - 200$ km/s,
corresponding to the outflow of lower-mass halos which are more numerous
in hierarchical structure formation. 
Considering all the gas particles, the no-wind (NW) run has 
slightly higher mass fraction at $v \leq 300$ km/s (not distinguishable in the plot), 
while runs with wind (CW, RVWa, RVWb) have higher gas mass with $v \geq 300$ km/s.


\begin{figure*} 
\centering 
\includegraphics[width = 0.95 \linewidth]{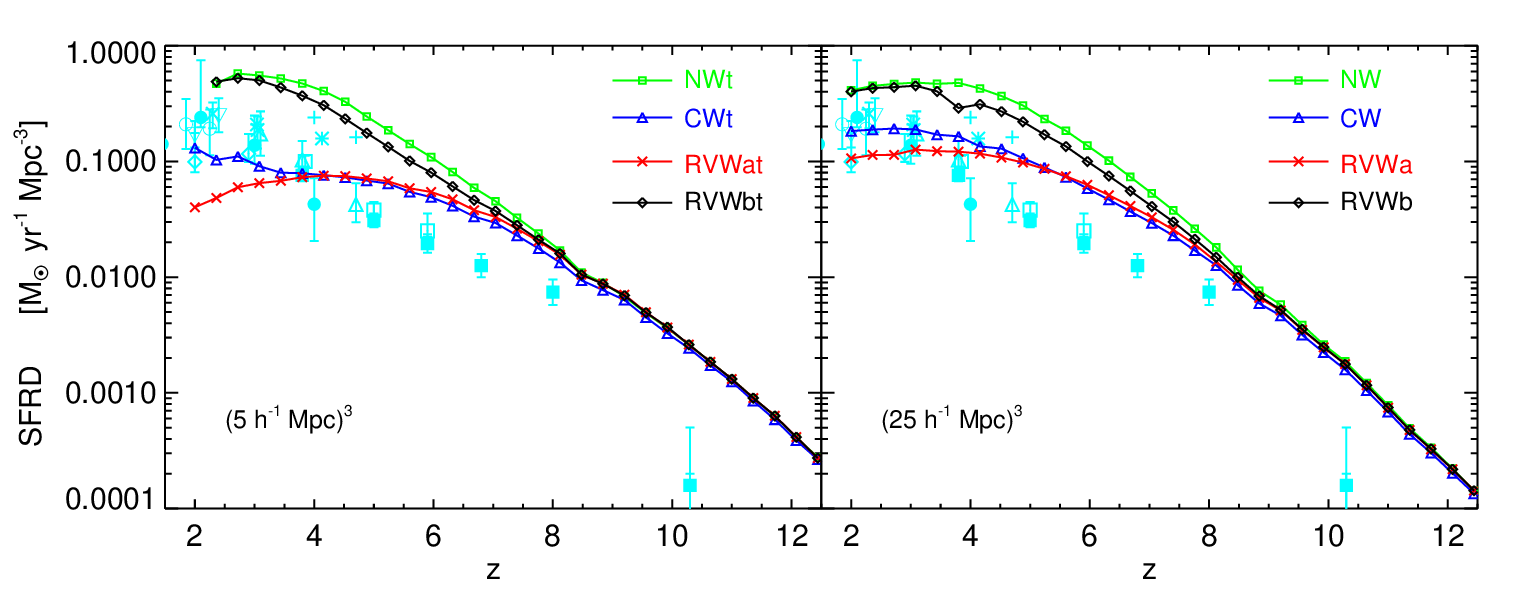} 
\caption{ 
Star formation rate density in whole simulation volume as a function of redshift, 
for the SB runs in the left and LB in the right panel, 
with the respective wind models labeled by the color and plotting symbol. 
The cyan symbols and error bars denote observational data from 
\citet{Cucciati12} - {\it filled circles}, and the compilations therein originally from 
\citet{Steidel99} - {\it asterisks}, 
\citet{Ouchi04} - {\it plus signs}, 
\citet{PerezGonzalez05} - {\it inverted triangles}, 
\citet{Schiminovich05} - {\it diamonds}, 
\citet{Bouwens09} - {\it open squares}, 
\citet{Reddy09} - {\it crosses}, 
\citet{Rodighiero10} - {\it open circles}, 
\citet{vanderBurg10} - {\it upright triangles}, 
\citet{Bouwens12} - {\it filled squares}. Detailed comparison is in \S\ref{sec-res-SFRD}. 
} 
\label{fig-SFRD} 
\end{figure*} 


\subsection{Star Formation Rate} 

\subsubsection{Global Star Formation Rate Density} 
\label{sec-res-SFRD} 

The global star formation rate density (SFRD) as a function of
redshift is plotted in Fig.~\ref{fig-SFRD}, 
for the SB runs in the left panel, and 
for the LB runs in the right one; in each panel 
the four wind models are labeled by the different colors and plotting symbols. 
The SFRD is computed by summing over all the SF occurring
in the whole simulation box and dividing it by the time-step interval
and box volume to obtain the rate density in $M_{\odot}$ yr$^{-1}$
Mpc$^{-3}$.  Galactic wind feedback clearly has significant
impact on SFRD, reducing SF several times depending on the outflow
model.

Between $z \sim 8 - 10$, there is already a trend of suppression of
SFRD with winds as the CW run has $1.2$ times lower SFRD than NW,
while at $z < 8$ the SFRDs of the runs diverge more. 
The differences are small between runs RVWb and NW; 
in RVWb the SFRD values 
are $1.5 - 2$ times smaller than NW between $z \approx 3 - 8$, and 
agree with NW at $z \approx 2 - 3$, making RVWb the wind model 
least effective in suppressing SF.  There is however a substantial
reduction of SFRD in runs CW and RVWa. The SB box 
have a stronger wind-driven suppression; SFRD is $12$ times smaller in
RVWat (and $5$ times smaller in CWt) than NWt at $z < 3$. 
While in the LB box, 
the reduction is $4$ times in RVWa and $2$ times in CW at $z < 4$. 
Analyzing the effective wind models, RVWa has $1.2$ times higher SFRD 
than CW from high $z$ up to $z = 5$, but later
RVWa suppresses it more and produces $2 - 4$ times lower SFRD at $z = 2$. 

Effects of box-size and resolution are visible in our results. 
The $z = 12$ SFRD is $\sim 2$ higher in the SB box than LB, 
because the SB series has higher resolution, and can track denser gas, forming more stars. 
However at $z < 3 - 4$, the RVWat run of SB series produces $2 - 3$ times lower SFRD than RVWa of LB, 
one reason for which is the relevant box-size; 
the halos forming in box SB are less-massive and fewer in number than in LB. 
Studying the shape of the SFRD evolution, 
the SB runs tend to show a peak in SFRD at a certain redshift: 
$z = 2.5$ in NWt and RVWbt, $z \leq 2$ in CWt, and $z = 4$ in RVWat. 
Whereas all the LB runs have a plateau of SFRD between $z = 2 - 3.5$. 

Observational data are overplotted in Fig.~\ref{fig-SFRD} with the cyan symbols and error bars. 
Each data set is shown with a different plotting symbol as listed next. 
These data are taken mainly from \citet{Cucciati12} - {\it filled circles}, 
and the compilations therein originally from 
\citet{Steidel99} - {\it asterisks}, 
\citet{Ouchi04} - {\it plus signs}, 
\citet{PerezGonzalez05} - {\it inverted triangles}, 
\citet{Schiminovich05} - {\it diamonds}, 
\citet{Bouwens09} - {\it open squares}, 
\citet{Reddy09} - {\it crosses}, 
\citet{Rodighiero10} - {\it open circles}, 
\citet{vanderBurg10} - {\it upright triangles}, 
\citet{Bouwens12} - {\it filled squares}. 

Comparing with observations, we see that no single simulation model
can fit the data from a single observation.  Taken collectively, there
is a better match of simulations with observations at low-$z$.  At
earlier cosmic epochs between $z \sim 4.5 - 10$, SFRD in the simulations 
is systematically higher, reaching $2 - 10$ times the observed values. 
At later times between $z \sim 2 - 4.5$, most of the 
observations lie within the ranges of SFRD produced by the different
wind models.  Run CW of the LB series is the model
providing the best-fit to the observations at lower-$z$.


\begin{figure*} 
\centering 
\includegraphics[width = 0.95 \linewidth]{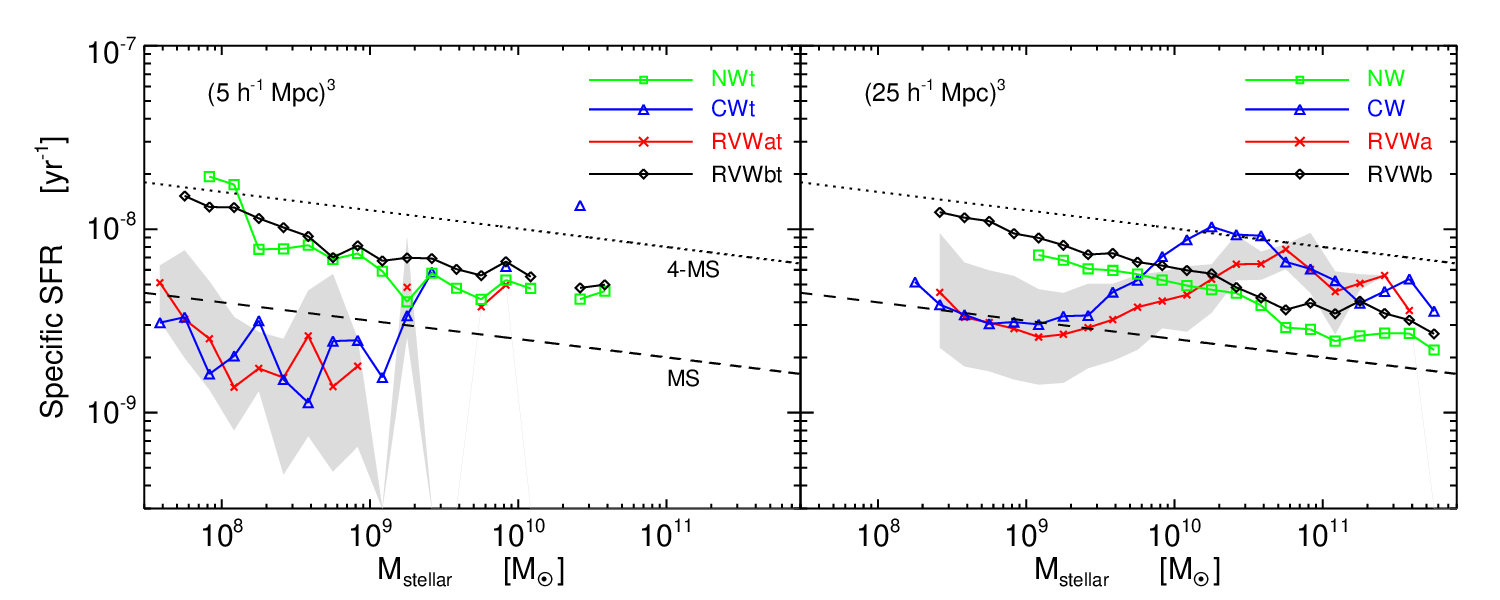} 
\caption{ 
Specific star formation rate as a function of galaxy stellar mass at $z = 1.98$, 
for the SB runs in the left and LB in the right panel, 
with the respective wind models labeled by the color and plotting symbol. 
The solid curves denote the median value within a mass bin for each run, 
and the grey shaded area enclose the $70$ percentiles above and below the median 
in runs RVWat and RVWa (red curves) showing the typical scatter. 
The black dashed line is observational data at $z = 2$ from \citet{Daddi07}, 
indicating the main sequence (MS) for star forming galaxies, 
SFR $= 200 \left( \frac{M_{\rm stellar}}{10^{11} M_{\odot}} \right)^{0.9}  M_{\odot}$ yr$^{-1}$. 
The black dotted line marks the loci 4 times above the MS; the region between 
MS and 4-MS encompass a majority fraction of observed galaxies from \citet{Rodighiero11}. 
} 
\label{fig-sSFR-vs-Mstar} 
\end{figure*} 


\subsubsection{Specific Star Formation Rate} 
\label{sec-res-sSFR} 

The specific star formation rate (sSFR) as a function of galaxy stellar mass at $z = 1.98$ 
is plotted in Fig.~\ref{fig-sSFR-vs-Mstar}. 
The SB runs are in the left panel and the LB runs in the right one; 
in each panel the four wind models are labeled by the different colors and plotting symbols. 
The solid curves denote the median value within a mass bin for each run, 
and the grey shaded area enclose the $70$ percentiles above and below the median 
in runs RVWat and RVWa (red curves) showing the typical scatter. 
The black dashed line is observational data at $z = 2$ from \citet{Daddi07}, 
indicating the main sequence (MS) for star forming galaxies, 
which is parametrized by the form: 
SFR $= 200 \left( \frac{M_{\rm stellar}}{10^{11} M_{\odot}} \right)^{0.9}  M_{\odot}$ yr$^{-1}$. 
The black dotted line marks the loci 4 times above the MS. 
According to \citet{Rodighiero11}, 
the region between MS and 4-MS encompass a majority fraction of observed galaxies. 

Comparing simulation results with observations, we see reasonably good agreement. 
The median sSFR of all the  models in the LB volume (right panel in Fig.~\ref{fig-sSFR-vs-Mstar}) 
lie between the observed MS and 4-MS. 
In the SB volume (left panel), runs NWt and RVWbt lie within MS and 4-MS, 
however CWt and RVWat are below MS, for a given stellar mass.

\subsection{Galaxy Mass Function and Mass Fraction} 
\label{sec-res-Mass-Fn-Frac} 

\begin{figure*} 
\centering  
\includegraphics[width = 0.9 \linewidth]{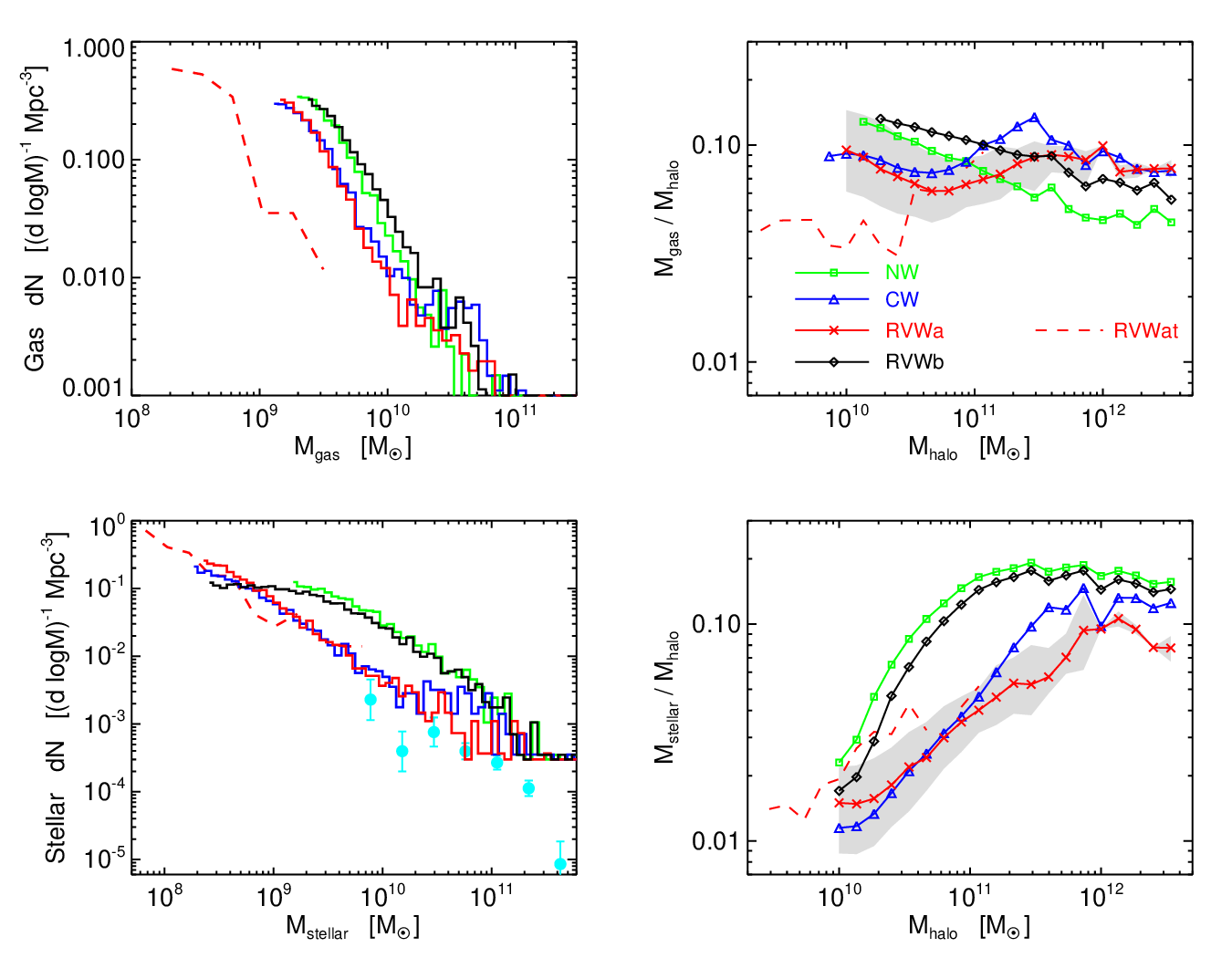} 
\caption{ 
Galaxy gas mass function (top-left), stellar mass function (bottom-left), 
along with gas mass fraction (top-right) and stellar mass fraction (bottom-right) w.r.t. total mass of halos, 
at $z = 2.23$, of the LB runs shown as the solid curves.   
In the right panels the solid curves denote the median value within a mass bin for each run, 
and the grey shaded area enclose the $70$ percentiles above and below the median 
in run RVWa (red curve) showing the typical scatter. 
One of the wind model (RVWat) from the SB box is plotted as the red dashed curve in each panel. 
The cyan filled circles and error bars in the bottom-left panel represent observational data 
of the stellar mass function from \citet{Marchesini09}, 
within $2 \leq z < 3$ (the results which are tabulated in their Table 1). 
Details are given in the text in \S\ref{sec-res-Mass-Fn-Frac}. 
} 
\label{fig-MassFn-MassFrac} 
\end{figure*}

The gas and stellar mass function ($dN$, in units of $(d \log M)^{-1}$ Mpc$^{-3}$) of galaxies 
in the LB runs at $z = 2.23$ is plotted as the solid curves 
in Fig.~\ref{fig-MassFn-MassFrac}, left panels. 
The top-left panel shows the mass function of the gas component in galaxies. 
All the 4 runs have the same slope between $M_{\rm gas} \sim (10^{9} - 10^{11}) M_{\odot}$, 
but shifted sideways as described next. 
NW and RVWb produce almost similar trends: 
$dN \approx 0.3$ at $(2 - 3) \times 10^{9} M_{\odot}$, 
with the RVWb mass function shifted right-ward by $\sim 1.2$ mass units $(M_{\odot})$. 
Results in CW and RVWa are quite similar, shifted left-ward with respect to NW by $(2 - 3) M_{\odot}$. 
The wind in these runs expel gas efficiently out of galaxies with $M_{\rm gas} \geq 10^{9} M_{\odot}$, 
producing a smaller number of objects with such high gas masses. 
Only at $\geq 10^{10} M_{\odot}$, CW has slightly more galaxies at the same mass than RVWa. 


The bottom-left panel of Fig.~\ref{fig-MassFn-MassFrac} shows the 
mass function of the galactic stellar component. 
Runs CW and RVWa having 
$dN \approx 0.2 - 0.3$ at $M_{\rm stellar} = (1.5 - 2) \times 10^{8} M_{\odot}$, 
produce a steeper stellar mass function than NW and RVWb where $dN \approx 0.1$ at $10^{9} M_{\odot}$. 
There is a small excess of galaxies in RVWa compared to CW between 
$M_{\rm stellar} = (2 \times 10^{8} - 10^{9}) M_{\odot}$, 
and a slight excess in NW than RVWb in the range $(10^{9} - 10^{10}) M_{\odot}$. 
The cyan filled circles and error bars represent observational data 
of the stellar mass function from \citet{Marchesini09}, 
within $2 \leq z < 3$ (the results which are tabulated in their Table 1). 
Our model RVWa (red curve) provide a reasonably good match to these high-$z$ observations 
over the mass range $M_{\rm stellar} = (8 \times 10^{9} - 10^{11}) M_{\odot}$. 
However our stellar mass functions are steeper than that of semi-analytic models by \citet{Bower12} 
at $z = 0$, and the observational data therein, 
along with other low-$z$ observations \citep[e.g.,][]{Papastergis12}. 

Fig.~\ref{fig-MassFn-MassFrac} also shows results of the wind model RVWat from the 
SB box, plotted as the red dashed curve in each panel. 
Comparing RVWa and RVWat, the smaller higher-resolution box has lower-mass halos forming, 
extending the gas mass function to $M_{\rm gas} \sim 10^{8} M_{\odot}$, 
and the stellar mass function down to $M_{\rm stellar} \sim 4 \times 10^{7} M_{\odot}$. 

The mass fractions of gas and stars in galaxies with respect to total mass of halos 
are plotted in the right panels of Fig.~\ref{fig-MassFn-MassFrac}: 
gas mass fraction in the top-right, and stellar mass fraction in the bottom-right. 
Solid curves denote the median value within a mass bin for each run 
labeled by the color and plotting symbol. 
The grey shaded area encloses the $70$th percentiles above and below the median 
in run RVWa (red curve), and show the typical scatter at given halo mass. 
With no-wind (NW) the gas fraction in galaxies decrease from 
$M_{\rm gas} / M_{\rm halo} \approx 0.13$ at $M_{\rm halo} = 10^{10} M_{\odot}$ 
to $0.04$ at $4 \times 10^{12} M_{\odot}$. 
The wind (CW, RVWa, RVWb) flattens the $M_{\rm gas} / M_{\rm halo}$ trend 
making it oscillatory around $(0.08 - 0.1)$ over the same $M_{\rm halo}$ range. 
The gas fractions in galaxies are lower than the cosmic baryon to DM fraction 
used in our simulations ($\Omega_{B, 0} / \Omega_{M, 0} = 0.17$). 
The gas mass fractions we obtain are roughly consistent with those 
found in massive cluster halos ($M_{500} > 10^{13} M_{\odot}$) at $z = 0$, 
in simulations including baryonic physics of cooling, SF and AGN feedback \citep{Planelles12}. 

The stellar fraction is largest in NW, 
because the SF rate is higher in the absence of wind (\S\ref{sec-res-SFRD}) 
producing more stars for the same halo mass. 
The wind in the other runs expel some star-forming gas out of galaxies, 
reducing the SF and consequently the stellar mass. 
$M_{\rm stellar} / M_{\rm halo}$ increases from $\approx (0.01 - 0.025)$ at $M_{\rm halo} = 10^{10} M_{\odot}$ 
to $\approx (0.08 - 0.16)$ at $6 \times 10^{11} M_{\odot}$, at a varying rate for the different runs, 
and remain almost flat at higher halo masses up to $4 \times 10^{12} M_{\odot}$. 
NW and RVWb have the steepest increase, followed by CW, and RVWa is the flattest. 
Observations also find the stellar mass fractions of low-$z$ galaxies 
increasing with halo mass between $M_{\rm halo} \sim (10^{10} - 10^{12}) M_{\odot}$ 
\citep[e.g.,][]{Papastergis12}, with a comparable slope, 
but the absolute values in observations are $\sim 10$ times smaller than our simulations. 

The gas and stellar fractions reveal the mass-dependence of the feedback 
acting on the galaxies in the different wind models. 
In runs NW and RVWb, more-massive halos have lower gas fraction than stars 
because of efficient conversion of gas to stars, 
and less-massive halos have more gas than stars; 
causing $M_{\rm gas} / M_{\rm halo}$ decrease with increasing $M_{\rm halo}$. 
In the effective wind models CW and RVWa, 
baryons are expelled efficiently and can escape from low-mass halos, 
causing the gas fraction (and star fraction) to be lower than NW and RVWb. 
At the high-mass end, outflows are not efficient to expel gas from halos, 
causing a higher gas fraction compared to cases NW and RVWb. 
The result is a flatter $M_{\rm gas} / M_{\rm halo}$ fraction versus $M_{\rm halo}$ 
in the effective outflow models. 
However the wind feedback affects the galaxy SFR; 
it is efficient enough and increases the gas cooling time, 
preventing the gas from turning into stars, 
causing the stellar fraction to be lower than NW and RVWb in massive halos. 

We thus conclude that the outflow models CW and RVWa have significant 
impact on the gas and stellar mass functions, as well as on the mass 
fractions of gas and stars in halos.


\begin{figure*}  
\centering 
\includegraphics[width = 1.0 \linewidth]{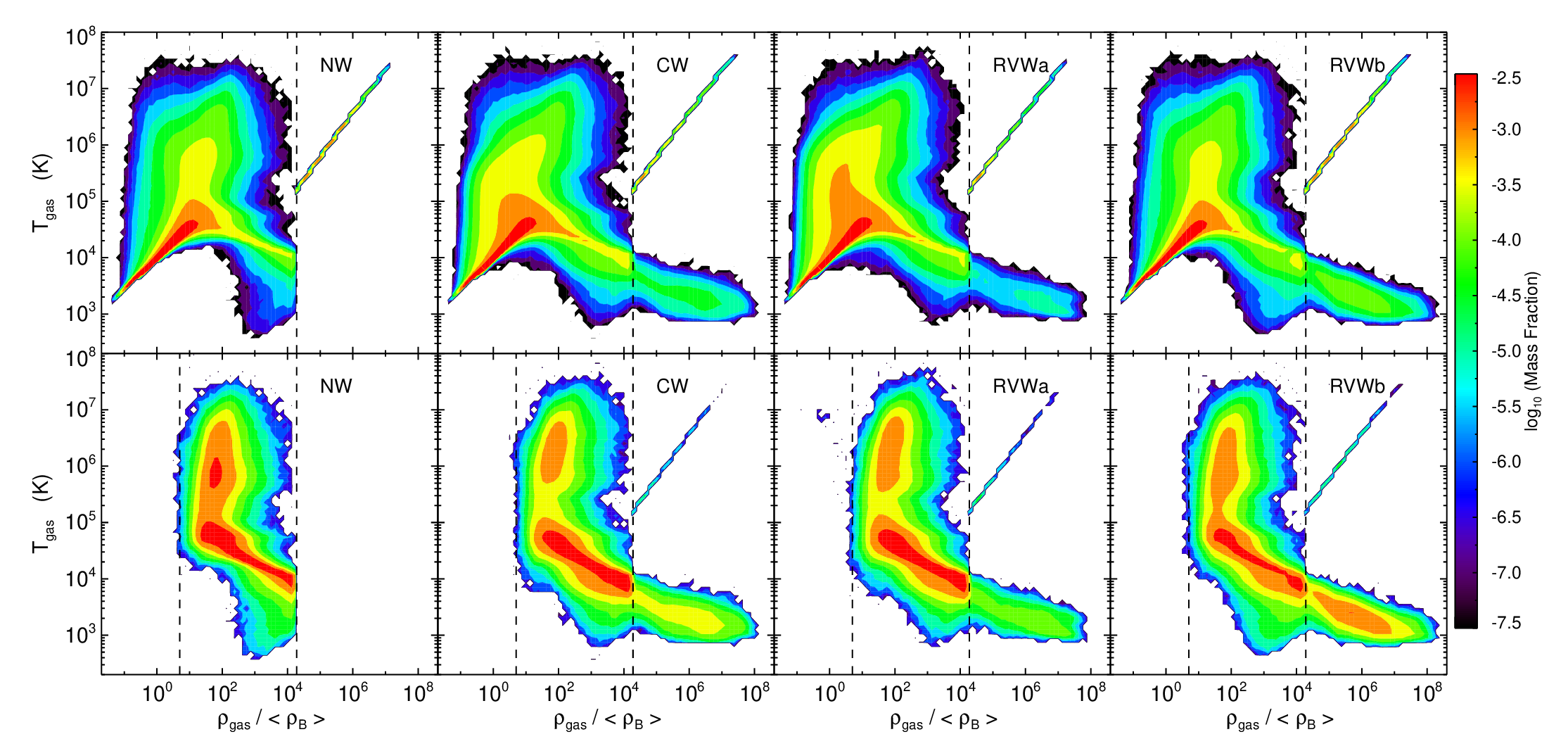} 
\caption{ 
Temperature versus density contrast (gas comoving density as a fraction of the mean baryon density) 
phase diagram of all the gas (top row) and the CGM gas (bottom row), at $z = 1.98$, 
for the LB runs with name labeled in each panel. 
The gas representing the CGM shown at the bottom is selected by tracking particles lying inside 
$R_{200}$ of all the halos and excluding the star-forming ones, which 
comprise of a number fraction (0.07 - 0.1) of all the gas particles shown in the corresponding top panel. 
The color code represents the mass fraction (in log-scale) of gas in density - temperature bins, 
with red showing the highest fractions going to black for the lowest. 
The single vertical dashed line in the top panels is the star formation threshold density 
(physical $n_{\rm SF} \approx 0.1$ cm$^{-3}$, \S\ref{sec-num-Subgrid}) 
above which a fixed equation of state is imposed on the gas in the multiphase SF model. 
The hot-phase temperature is shown for the multiphase gas particles, 
which results in the constant-slope straight line above $n_{\rm SF}$ 
in the top-right portion of each panel (\S\ref{sec-res-T-rho}). 
The two vertical dashed lines in the bottom panels represent the bounding densities of the CGM: 
the star formation threshold density is the upper limit, 
and a lower limiting density $\delta_{\rm CGM} = 5$ inferred from this plot. 
} 
\label{fig-Rho-vs-T} 
\end{figure*} 


\subsection{Temperature-Density Phase Diagram} 
\label{sec-res-T-rho} 

The temperature versus density phase diagram of all the gas at $z = 1.98$ 
is plotted in Fig.~\ref{fig-Rho-vs-T}, top row, for the LB runs in 4 horizontal panels, 
color coded by the gas mass fraction. 
The larger boxsize is chosen for the two $[T - \rho]$ diagrams we present in this section, 
because it gives a statistics over a larger range of halos. 
The $x$-axis denotes the density contrast, 
$\delta = \rho_{\rm gas} / \langle \rho_B \rangle$, 
which is the comoving density of gas $\rho_{\rm gas}$ as a fraction of the 
comoving mean baryon density, $\langle \rho_B \rangle = \Omega_{B, 0} 3 H_0^2 / (8 \pi G)$. 
The vertical dashed line is the SF threshold density 
($n_{\rm SF}$, \S\ref{sec-num-Subgrid}) 
above which a fixed effective equation of state is imposed on the gas 
in the multiphase SF model (SH03, \citealt{Wiersma09b}). 
The hot-phase temperature is shown for the multiphase gas particles, 
which forms the constant-slope straight line above $n_{\rm SF}$ 
in the top-right portion of each run. 
In the NW run (top-left panel), all the gas denser than $n_{\rm SF}$ 
follows the equation of state, and there is no cold gas at those high densities. 

In the runs with wind, a fraction of the gas denser than $n_{\rm SF}$ 
is cooler ($< 10^4$ K) than the NW case, forming a 
cold, dense tail in the bottom-right portion of the phase diagram. 
This is a consequence of the decoupling formalism of the galactic
outflow implementation (\S\ref{sec-num-Implement}).  The wind
particles are allowed to move away from SF regions and no longer follow
the SF effective equation of state.  They are however still very
dense, where the cooling rate is high and cooling time short.
Therefore they cool very fast, soon after being kicked into wind,
and form the cold, dense extension above $n_{\rm SF}$ in the phase
diagram.  Subsequently, if the kick velocity is higher than the
neighbors, the wind particles exit the dense SF region soon, moving to
lower-density ($< n_{\rm SF}$), hotter ($> 10^4 - 10^5$ K) regions.
Thus in our galactic outflow models, the wind is launched from dense
SF phase, and soon after that the wind goes through a cold phase.

The remaining of the phase diagram looks qualitatively similar in the 4 runs. 
The majority of the gas (red in Fig.~\ref{fig-Rho-vs-T}) lies in 
underdense to moderately overdense ($\delta \sim 10^{-1} - 10^{2}$) 
and cool-warm ($T \sim 10^3 - 10^5$ K) phase, which is the 
cosmic baryons in balance between adiabatic cooling and photoionization background heating. 
A small fraction of gas at a higher range of overdensities ($\delta \sim 10^{-1} - 10^{4}$) 
is hot ($T \sim 10^6 - 10^8$ K), being heated by shocks during gravitational collapse. 
At $\delta \sim 10^{2} - 10^{4}$, there is another 
cold ($T < 10^5$ K) phase which is the dense gas cooling into DM halos to form galaxies. 

The gas mass fraction in the phases varies between the 4 runs, 
forming different patterns in the phase diagram. 
Runs NW and RVWb have a higher fraction of gas in the multiphase SF branch 
of effective equation of state. 
Among the wind runs, RVWb has the highest mass fraction in the cold wind phase, 
followed by CW and finally RVWa. 
This is because of the different wind kick velocity in each case; 
RVWb has the lowest $v_w$ (seen in Fig.~\ref{fig-VelGas_vs_Dist}, \S\ref{sec-res-OutVel}) 
where gas is not able to escape the dense phase, 
while in RVWa the wind particles are able to escape the dense phase quickly. 
Runs CW and RVWa have a higher fraction of underdense ($\delta < 1$), warm-hot ($T \sim 10^4 - 10^6$ K) 
gas, denoted by the red and yellow contours in Fig.~\ref{fig-Rho-vs-T}, 
composed of gas previously ejected from dense SF regions of galaxies as strong winds. 

The temperature - density phase diagram of only the CGM gas at $z = 1.98$ 
is plotted in Fig.~\ref{fig-Rho-vs-T}, bottom row, for the LB runs in 4 horizontal panels. 
Gas particles lying inside the virial radius $R_{200}$ 
(computed analytically using Eq. \ref{eq-Mhalo}) of all the FOF halos are selected, 
excluding the star-forming ones (ie. particles with SFR = 0 are only counted). 
The selected CGM gas comprise of a number fraction (0.07 - 0.1) of 
all the gas particles in a run, and is shown in Fig.~\ref{fig-Rho-vs-T}, 
color coded by the gas mass in $[T - \delta]$ bins as a fraction of the total mass plotted. 
The densities defining the boundaries of the CGM are denoted by the vertical dashed lines. 
The SF threshold density is the upper limit, 
above which gas forms stars and becomes part of galaxy. 
From Fig.~\ref{fig-Rho-vs-T}, bottom row, 
we infer a lower limiting density $\delta_{\rm CGM} = 5$ of the CGM. 

The NW run (bottom-left panel) has no CGM gas denser than $n_{\rm SF}$, 
because of imposing the SFR = 0 criterion for selecting CGM particles, 
as here all the gas denser than $n_{\rm SF}$ is star-forming. 
The runs with wind have a fraction of gas denser than $n_{\rm SF}$ which is 
not star-forming, and they are the wind-phase particles: 
mostly cold undergoing hydrodynamic decoupling, 
and a small fraction which recently received a wind kick and still in the hot multiphase branch. 
Noting the effect of feedback, 
we see that the CGM is separated into a cold $T < 10^5$ K phase and hotter gas 
in a more well-defined way in models CW and RVWa impacted by their effective outflows, 
than in NW and RVWb. 

Overall, analyzing the $[T - \delta]$ plane, we can conclude that the
thermal properties of the IGM remain roughly the same between the
different outflow models; with the winds producing a higher fraction 
of underdense, warm-hot gas. 
Furthermore the CGM is well separated into cold and warm-hot phases by the winds.

\subsection{Single Galaxy Gas Kinematics} 
\label{sec-res-SingGal}

\begin{figure*} 
\centering 
$ 
\begin{array}{c} 
\includegraphics[width = 1.0 \linewidth]{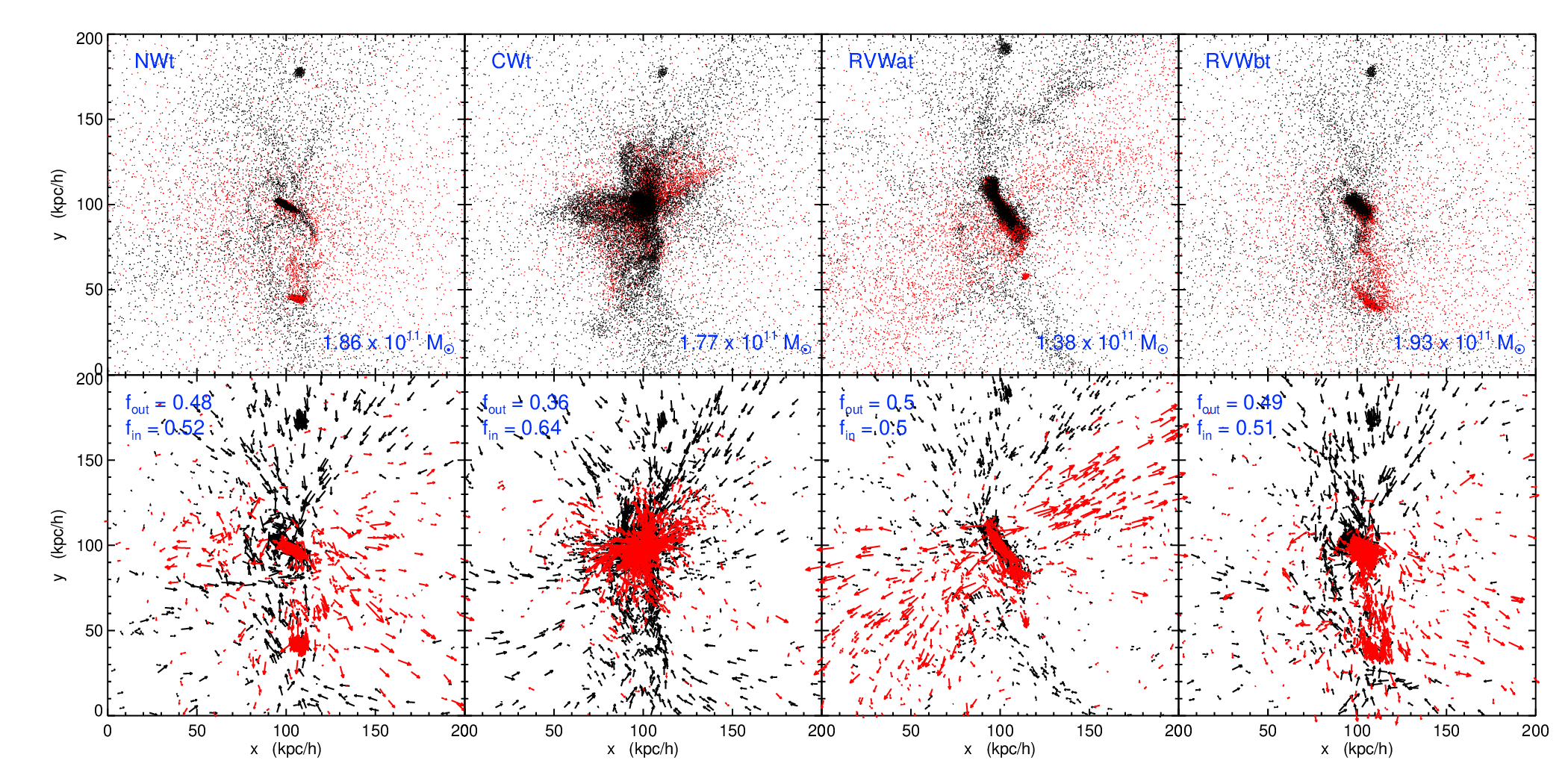} \\ 
\includegraphics[width = 1.0 \linewidth]{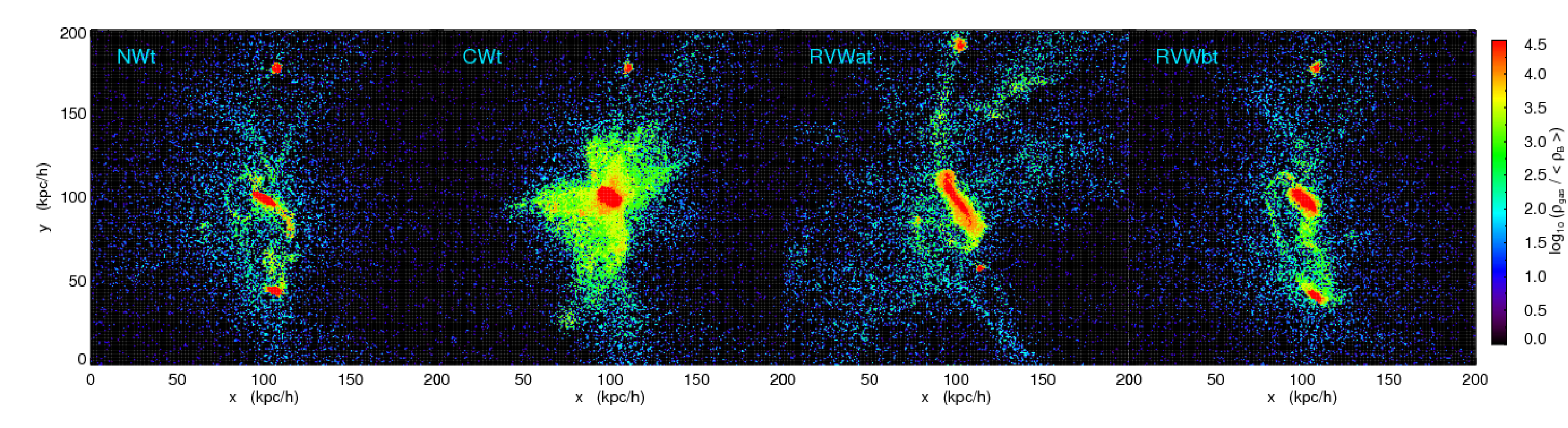} \\ 
\includegraphics[width = 1.0 \linewidth]{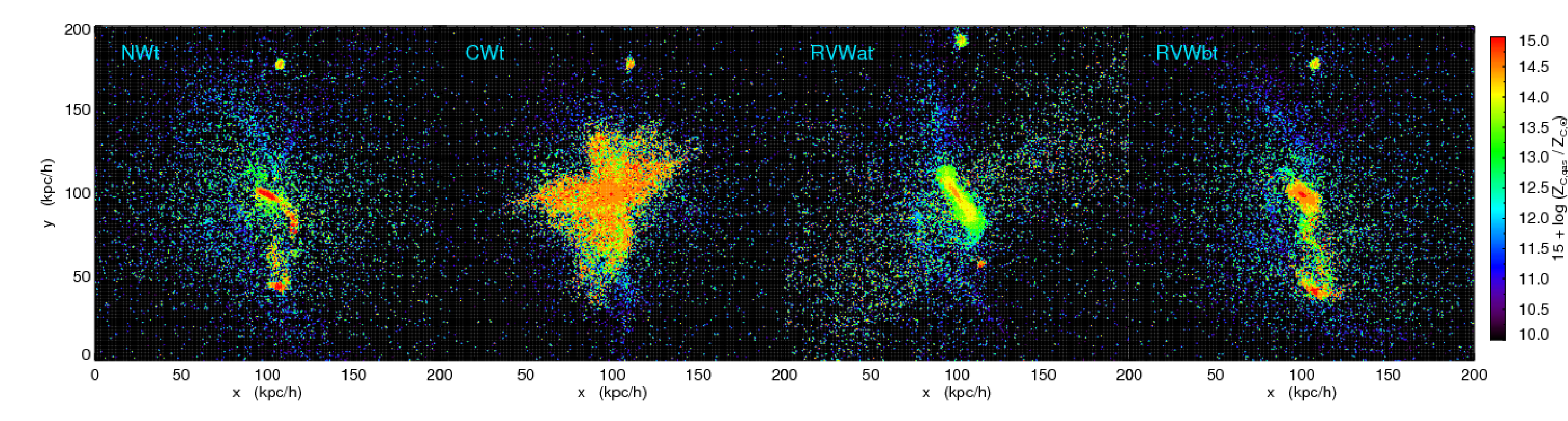} 
\end{array} 
$ 
\caption{ 
Projection of gas kinematics in the $[x - y]$ plane of a $(200 h^{-1}$ kpc$)^3$ volume 
centered around the maximum gas density location of the most-massive galaxy 
at redshift $z = 2.12$, 
in the SB simulations with different wind models (from the left): 
no wind, energy-driven constant-velocity, radially varying with fixed parameters, 
and that with halo mass dependent parameters. 
In the top two rows the outflowing ($v_r > 0$) particles are denoted as red, 
and the inflowing ($v_r < 0$) as black. 
First row from the top shows the positions of all the gas particles within the projected volume, 
and the second row depicts the velocity vectors of $10 \%$ of each component. 
The total halo mass of each galaxy, and the number fraction of outflowing ($f_{\rm out}$) 
and inflowing ($f_{\rm in}$) gas are written within the panels. 
The inflow is overplotted on the outflow in the first row, and vice versa in the second. 
Third row shows gas density contrast, and fourth row is carbon metallicity, 
color coded from red as the highest and black as the lowest values. 
} 
\label{fig-XYplane-Pos-Vel-rho-ZC} 
\end{figure*}

The most-massive galaxies at $z = 2.12$ in the SB simulations 
are plotted in Fig.~\ref{fig-XYplane-Pos-Vel-rho-ZC}. 
The smaller boxsize is used here again for ease of visualization 
(like in \S\ref{sec-res-OutVel}), to clearly distinguish the outflowing gas 
in the plot of a massive galaxy within the box volume. 
Each of the 4 rows shows a unique gas property, the different wind model runs 
are labelled in the top-left corner of the 1st, 3rd and 4th rows. 
The maximum gas density location is at the center, and the panels show 
the projection of gas kinematics in the $[x - y]$ plane 
of a surrounding $(200 h^{-1}$ kpc$)^3$ volume. 
The total halo mass (DM + gas + stars) of the galaxies are in the range 
$M_{\rm halo} = (1.4 - 1.9) \times 10^{11} M_{\odot}$. 
One can see the formation of a galaxy disk in runs NWt, RVWat and RVWbt of varying sizes, 
whereas in CWt the galaxy looks irregularly shaped in the projected plane. 
The galaxy disk in model RVWat is the largest in size, followed by NWt and then RVWbt. 
The disks in NWt and RVWbt are quite similar, with RVWbt producing a thicker disk. 


In the top two rows the outflowing (radial velocity with respect to galaxy center, $v_r > 0$) 
particles are denoted as red, and the inflowing ($v_r < 0$) as black. 
The first row shows the positions of all the gas particles within the projected volume, 
and the second row depicts the velocity vectors of 
$10 \%$ outflowing gas particles and $10 \%$ inflowing gas. 
The number fraction of outflowing ($f_{\rm out}$) and inflowing ($f_{\rm in}$) gas 
are written in the top-left  corner of the second row panels. 
Within this $(200 h^{-1}$ kpc$)^3$ projected volume, 
in runs NWt, RVWat and RVWbt, about half of the gas is inflowing and the other half outflowing. 
While in run CWt more gas $(64 \%)$ is undergoing infall than outflow $(36 \%)$. 
This is likely because of fall-back of gas which was kicked outward at earlier epochs, 
but did not escape the halo potential, and is infalling back. 
The in/out flow fractions in run CWt are similar to the result by \citet{Shen12} who found that 
in the {\it Eris2} zoom-in simulation, the CGM of a Milky Way-type galaxy at $z \sim 3$ 
has about one third of all the gas within the virial radius as outflowing. 
In our Fig.~\ref{fig-XYplane-Pos-Vel-rho-ZC}, 
the inflow is overplotted on the outflow in the first row 
depicting a prominence of black points over red at the center, 
and vice versa in the second row showing a prominence of red points over black; 
this demonstrates the mixture of gas dynamics 
(i.e. contains both inflow and outflow) in the central $20 - 30$ kpc regions. 

The outflow is not well-structured in the no-wind (NWt) and  
the low-velocity wind model (RVWbt) cases, some gas escaping 
along the direction of least resistance through a low-density void in the bottom-right half. 
In run CWt most of the outflowing gas lies inside $r < 50$ kpc, because here 
the wind kick velocity $v_w = 400$ km/s is not enough to drive the outflow to larger distances; 
the strong gravitational potential of the halo causes the gas to fall back inward 
also resulting in a higher inflow fraction in this run. 
The radially-varying subgrid wind velocity $v_w(r)$ of Eq.~\ref{eq-vSteidel} in run RVWat 
produces a well-developed gas outflow propagating perpendicular to the galaxy disk, 
escaping to $r > 100$ kpc from the center, 
seen as the red arrows along top-right and bottom-left in the 3rd panel of the 2nd row. 

The third row of Fig.~\ref{fig-XYplane-Pos-Vel-rho-ZC} 
shows the gas density contrast 
color coded on a log-scale from red as the highest and black as the lowest values. 
All the runs have a central overdense region, where SF occurs forming a  galaxy disk in some cases. 
Wind particles are kicked from these SF regions, 
carrying a fraction of the central gas to larger distances. 
In run CWt the constant-velocity wind forms an extended halo of gas at $r \leq 50$ kpc, 
seen as yellow-green in the 2nd panel, 
which is most likely bound to the galaxy potential and not able to escape. 
The strong radially-varying wind in RVWat produces a more extended ($r \sim 100$ kpc) 
but lower-density diffuse outflow, which is likely escaping the galaxy in the 3rd panel. 
The low-velocity wind run RVWbt has most of the gas concentrated in the central $10 - 20$ kpc. 

Carbon metallicity ($Z_C$, defined in \S\ref{sec-num-sim}) is plotted in the 
fourth row of Fig.~\ref{fig-XYplane-Pos-Vel-rho-ZC}. 
Runs NWt and RVWbt have a larger central concentration of metals 
(red in the figure) originating from SF, 
because there is either no-wind or the wind is not effective to spread the metals around. 
The central metal content is smaller in runs CWt and RVWat, 
since winds carry the metals out from the SF regions and enrich the CGM and IGM. 
Consequently, in the 2nd and 3rd panels, the distribution of metals outside galaxy disk 
is aligned with the outflowing velocity (red arrows in the 2nd row), 
and two extended gas outflow regions (density contrast in the 3rd row) are present.

The large scale structure of all the runs are similar as expected from the same initial condition, 
though there are small-scale differences as described above arising from different wind models. 
There is gas inflow from bottom / bottom-right regions of the panels into the center 
and from top / top-right into the center, 
seen as black points in the 1st row, black inflowing arrows in the 2nd, 
overdense filaments in the 3rd, which have lower-metallicity as seen in the bottom row. 
These likely denote pristine gas infall along cosmological filaments 
that ``feed'' the galaxy at the center. 
Another common feature is the overdense gas clump $70 - 80$ kpc to the top of the center, 
most likely an infalling sub-structure which will eventually merge with the central galaxy.

\subsection{Radial Profiles Around Galaxy Centers at $z \sim 2$} 
\label{sec-res-Radial-Profile} 

\begin{figure*} 
\centering 
\includegraphics[width = 1.0 \linewidth]{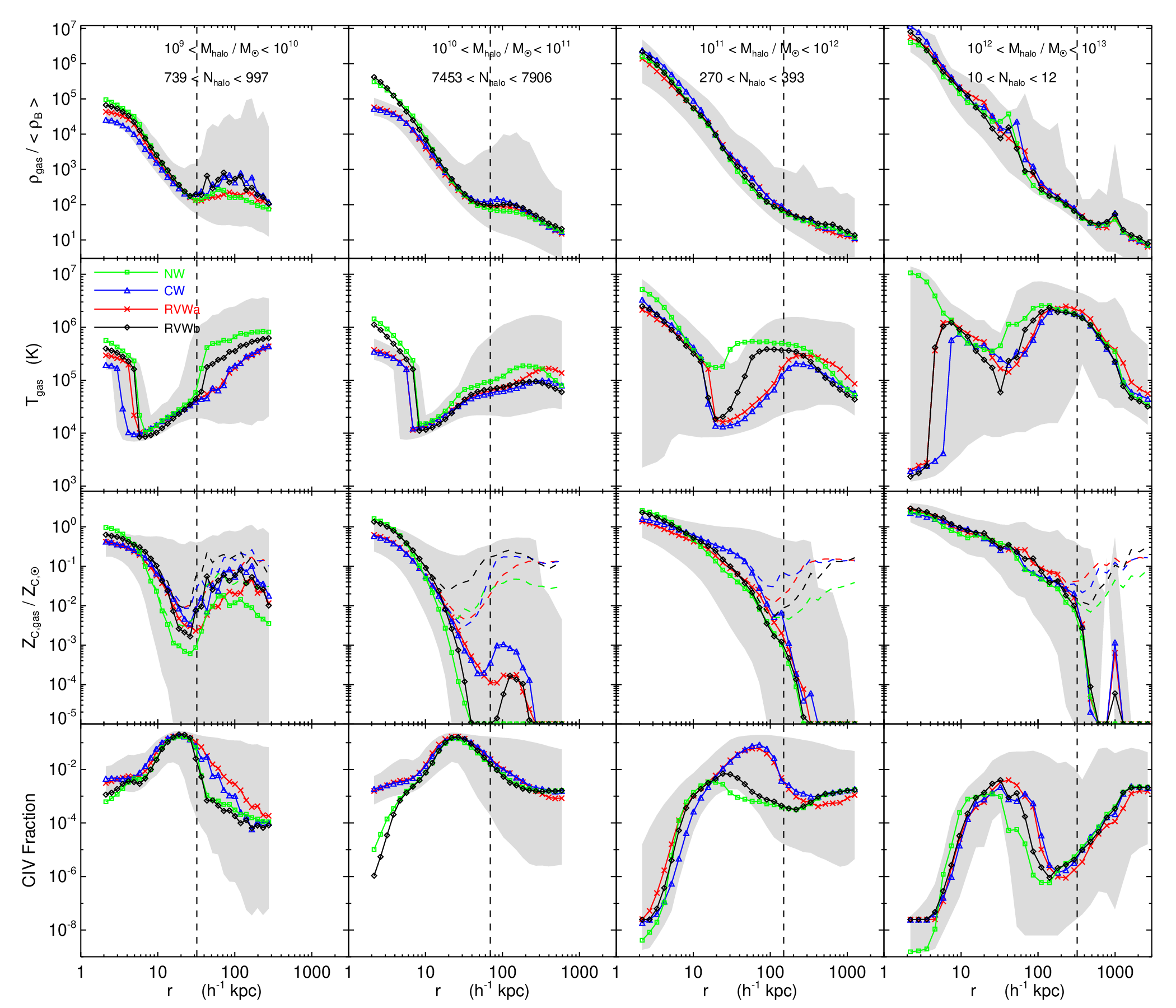} 
\caption{ 
Radial gas profiles of galaxies found by FOF group finder 
at $z = 1.98$. 
Each row shows a gas property (described later) as a function of comoving radius 
(or distance from the position of maximum density considered as the galaxy center), 
for four total (DM + gas + star) halo mass ($M_{\rm halo} / M_{\odot}$) ranges 
in the 4 horizontal panels: 
$10^{9} - 10^{10}$ (left), $10^{10} - 10^{11}$ (second from left), 
$10^{11} - 10^{12}$ (third from left), and $10^{12} - 10^{13}$ (right). 
All the halos within each mass range (the number mentioned in the top row panels) 
are stacked over for each run, and the plotted curve denotes the median value in a radial bin. 
The grey shaded area enclose the $70$th percentiles 
above and below the median in run RVWa (red curve), showing the typical radial scatter. 
The vertical dashed line is the analytical expression for the 
halo virial radius $R_{200}$ in comoving coordinates (from Eq. \ref{eq-Mhalo}), 
for $M_{\rm halo} / M_{\odot} = 3 \times 10^{9}$, $3 \times 10^{10}$, 
$3 \times 10^{11}$, and $3 \times 10^{12}$ in the panels from left respectively in each row. 
First (top-most) row is the gas density contrast. 
Second row is gas temperature, 
and hot-phase temperature is shown for the multiphase gas particles (\S\ref{sec-res-Radial-T}). 
Third row is carbon metallicity, 
showing the ratio of carbon mass fraction in gas to that of the Sun 
($Z_{C, \odot} = 0.002177$).        
The dashed curves in the third row represent median-$Z_C$ for the enriched particles only 
in each radial bin, i.e. not counting particles with zero-$Z_C$ (\S\ref{sec-res-Radial-ZC}). 
Fourth row is the fraction of CIV (triply ionized carbon) in the gas    
(computed from simulation redshift, gas density, temperature 
and assuming a photoionizing background using CLOUDY ionization tables). 
} 
\label{fig-rho-T-ZC-fC4-vs-R-MassBins} 
\end{figure*}

The radial profiles of gas properties around centers of galaxies
(found by FOF group finder) at $z = 1.98$ 
for the LB runs are plotted in Fig.~\ref{fig-rho-T-ZC-fC4-vs-R-MassBins}. 
The larger boxsize is chosen for all the radial profiles we present in this section, 
since more number of halos and more massive halos form in the larger box; 
it hence gives a wider statistics. 

Each row of Fig.~\ref{fig-rho-T-ZC-fC4-vs-R-MassBins} shows a property
as a function of comoving radius (or distance from the maximum density
position considered as the galaxy center), by counting all the gas
particles lying inside a distance $R_{\rm lim}$ from the center, 
in the following format. 
The four horizontal panels denote total (DM + gas + star) halo mass 
($M_{\rm halo} / M_{\odot}$) ranges: 
$10^{9} - 10^{10}$ (left) with number of halos in the four plotted runs 
within the range $N_{\rm halo} = 739 - 997$; 
$10^{10} - 10^{11}$ (second from left) having $N_{\rm halo} = 7453 - 7906$; 
$10^{11} - 10^{12}$ (third from left) where $N_{\rm halo} = 270 - 393$; 
$10^{12} - 10^{13}$ (right) with $N_{\rm halo} = 10 - 12$. 
All the halos within each mass range are stacked, and the plotted solid curve denotes 
the median value in radial bins for each run labeled by the color and plotting symbol. 
The grey shaded area encloses the $70$th percentiles above and below the median 
in run RVWa (red curve), showing the typical               
scatter at a given radius, 
since galaxies in general do not have spherically-symmetric properties. 
The vertical dashed line is the halo virial radius $R_{200}$ 
in comoving coordinates (analytical expression from Eq. \ref{eq-Mhalo}) 
for the following masses in the horizontal panels from left: 
$M_{\rm halo} / M_{\odot} = 3 \times 10^{9}$, $3 \times 10^{10}$, 
$3 \times 10^{11}$ and $3 \times 10^{12}$, 
where the exact values are $R_{200} = 32.2, 69.4, 149.6$, 
and $322.2 ~ h^{-1}$ kpc respectively. 
The outer plotting radius $R_{\rm lim}$ is chosen to be $300 ~ h^{-1}$ kpc for 
$M_{\rm halo} / M_{\odot} = 10^{9} - 10^{10}$. 
In the other halo mass ranges $R_{\rm lim}$ is scaled up according to the virial radius, 
making $R_{\rm lim} = 300, 646.6, 1393.8$, and $3001.9 ~ h^{-1}$ kpc. 


\subsubsection{Density} 
\label{sec-res-Radial-Density} 

The gas density contrast (ratio of comoving density to the comoving mean baryon density, 
\S\ref{sec-res-T-rho}) radial profiles are plotted in the top-most first row 
of Fig.~\ref{fig-rho-T-ZC-fC4-vs-R-MassBins}. 
Within the approximate virial radius $r < R_{200}$ all the profiles are composed of two 
negative-sloped functions separated by a threshold radius dependent on halo mass and wind model. 
At $r > R_{200}$ the density profiles 
tend to rise again and fall, forming a local peak. 
This occurs because of the presence of other smaller halos and substructures, 
giving rise to a local density peak. 
For each halo mass range  a sufficiently large volume 
within $r \sim 10 R_{200}$ is plotted to reach surrounding structures. 
This trend is most prominent in the lowest mass halos, and decreases at higher masses. 

The inner parts $r < 10 h^{-1}$ kpc of the 2 lower halo mass ranges 
($10^{9} - 10^{10}$ and $10^{10} - 10^{11}$, left two panels) present significant differences: 
CW and RVWa produce a lower density than the NW and RVWb cases, 
because the strong wind is able to expel gas from the star-forming regions, 
consequently reducing the central gas density by $10 - 30$ times. 
The trend is almost reversed in the outer parts ($r > R_{200}$): 
CW, RVWa and RVWb has a somewhat higher density than NW, because of accumulation of gas expelled by wind. 

The differences are smaller in the two higher halo mass ranges 
($10^{11} - 10^{12}$ and $10^{12} - 10^{13}$, right two panels), because the wind, 
not being dependent on halo mass in runs CW and RVWa, is less effective in ejecting gas out of massive galaxies. 
CW produces slightly higher density (by $2 - 3$ times) than RVWa in the inner $r < 10 h^{-1}$ kpc, 
while outside this distance they are quite similar. 
Comparing our density radial profiles in the halo mass range $10^{12} - 10^{13}$ 
(top-right panel in Fig.~\ref{fig-rho-T-ZC-fC4-vs-R-MassBins}), 
with that of \citet{Hummels12} Fig. 5, we find qualitative agreements; 
however our density profiles are steeper.

\subsubsection{Temperature} 
\label{sec-res-Radial-T} 

The temperature radial profiles are presented 
in the second row of Fig.~\ref{fig-rho-T-ZC-fC4-vs-R-MassBins}, 
where the hot-phase temperature has been used for those gas particles 
which are multiphase (star-forming). 
In the inner parts $r \leq 10 h^{-1}$ kpc of the galaxies, 
the $T$-profiles follow the negative-sloped density-profiles 
(first row of Fig.~\ref{fig-rho-T-ZC-fC4-vs-R-MassBins}), 
for all the runs in the left 3 panels, and run NW in the right panel. 
This represents the dense gas near galaxy center undergoing SF, 
and having a high temperature ($\sim 10^{5} - 10^{7}$ K) 
as a result of following the SF effective equation of state (\S\ref{sec-res-T-rho}). 
In the wind runs there is also the presence of cooler ($< 10^4$ K) gas, 
which has recently received a wind kick and 
undergoes hydrodynamic decoupling (\S\ref{sec-num-Implement}). 
This gas was seen as the cold, dense tail in the bottom-right portion of the 
$[T - \rho]$ phase diagram (Fig.~\ref{fig-Rho-vs-T}, \S\ref{sec-res-T-rho}). 
Consequently, runs CW, RVWa and RVWb 
present a bimodal $T$ distribution at $r \leq 10 h^{-1}$ kpc in some panels, 
composed of hot multiphase SF and cold wind. 

There is a change in $T$ slope in the outer parts, at $r \geq 6 h^{-1}$ kpc 
in the left 2 panels and at $r \geq (20 - 30) h^{-1}$ kpc in the right 2 panels, 
when the gas $T$ increases with radius, likely because of shock heating at galaxy outskirts. 
Between $(10 - 300) h^{-1}$ kpc, the NW model results in the highest-$T$ 
($\sim 10^{5} - 10^{6}$ K), followed by RVWb, then RVWa and CW. 
Our temperature profiles show a bump at $200 - 300$ kpc, 
which is at a larger-$r$ than the peak of \citet{Hummels12}. 


\subsubsection{Carbon Metallicity}          
\label{sec-res-Radial-ZC} 

Radial profiles of carbon metallicity are plotted in 
the third row of Fig.~\ref{fig-rho-T-ZC-fC4-vs-R-MassBins}, 
showing the ratio of carbon mass fraction in the gas to that of the Sun. 
The solid curve medians and percentile values are computed considering all 
(both enriched and non-enriched) gas particles in radial bins. 
The dashed curves represent median-$Z_C$ for the enriched particles only, 
i.e. those having $Z_C > 0$, without counting particles with $Z_C = 0$. 

We also checked the gas fraction (by particle number) which has $Z_C > 0$. 
All the gas is enriched inside a limiting radius, 
which is $\sim 7, 10, 40$, and $120 ~ h^{-1}$ kpc in the panels from left, 
defining the size of the central SF region or the sites of metal generation. 
The enriched fraction decreases outside the limiting radius, 
with a trend depending on halo mass and wind model: 
it reduces to $(0.5 - 0.8)$ at $R_{200}$ and $(0.45 - 0.7)$ at $300 ~ h^{-1}$ kpc. 
Runs CW and RVWa enrich a higher (up to $1.4$ times) fraction of gas than NW and RVWb. 

The dashed median $Z_C$ in Fig.~\ref{fig-rho-T-ZC-fC4-vs-R-MassBins}, third row 
is indistinguishable from the solid median $Z_C$ in the inner parts, 
since most of the gas is enriched. 
While at large $r$ the dashed median $Z_C$ is higher, because the 
contribution of the non-enriched ($Z_C = 0$) particles reduces the solid median $Z_C$, 
as can be seen in all the four panels. 
Especially in the second and third panels the enriched median $Z_C$'s are 
$100 - 5000$ times higher than the total median $Z_C$'s. 

Some features of $Z_C$ are similar to the 
gas density profiles (\S\ref{sec-res-Radial-Density}), 
because metals are produced during SF which occurs in dense regions. 
At $r < (0.4 - 0.6) R_{200}$ all the profiles show decreasing $Z_C$ going outward from center, 
with varying $r$-dependent negative slopes, which we later fit by a second order polynomial. 
The $Z_C$ profiles start to rise again from $r \geq (0.4 - 0.6) R_{200}$ and fall at larger $r$, 
forming a local peak at $r > R_{200}$. 
Such a trend is visible in the left two and right-most panels for both solid and dashed curves, 
and in the other panels for the dashed curves only. 
It occurs because of a combination of reasons: 
the presence of surrounding substructures where more metals are produced in-situ by ongoing SF, 
and the spreading of metals by wind from the central SF regions into the CGM. 
These simulation results are consistent with observations 
which show breaks (changes of slope) in the radial metallicity profiles, 
and/or rising metallicity gradients in the outer regions of galaxies \citep[e.g.,][]{Scarano12}. 

In the inner $r < (5 - 7) h^{-1}$ kpc CW and RVWa profiles have a lower $Z_C$ than NW and RVWb, 
because wind suppresses central SF and transports some metal out. 
The trend reverses in the outer $r > (5 - 7) h^{-1}$ kpc: 
CW, RVWa and RVWb produce a metallicity about $20 - 30$ times higher than NW, 
because of accumulation of metal-enriched gas expelled by wind. 
The differences are most prominent in the lower halo mass ranges 
($10^{9} - 10^{10}$ and $10^{10} - 10^{11}$, left two panels), and decreases at higher masses. 
Massive halos of $10^{11} - 10^{12}$ (third from left) still show 
significantly different profiles at large-$r$, 
while the profiles of the four wind models in the most-massive halos (right) are very similar. 

Model RVWb produces noteworthy changes in the $Z_C$ profile compared to NW: 
metallicity in the central part is slightly lower in RVWb reaching about half of NW, 
but starts to become larger at $r > 5 h^{-1}$ kpc up to $10$ times higher than NW. 
This shows that the wind in RVWb, though least effective in other aspects, 
is substantially effective in transporting metals away from SF regions into lower-density surroundings. 

We perform polynomial fits to the $Z_C$ radial profiles of runs NW and RVWa, 
within certain radial limits, for the 4 halo mass ranges plotted in Fig.~\ref{fig-rho-T-ZC-fC4-vs-R-MassBins}. 
The median $Z_C$ (solid curves) versus $r$ data are fitted with a second order polynomial of the form, 
$\log (Z_C / Z_{C, \odot}) = A + B (\log r) + C (\log r)^2$, 
and the resulting fit coefficients are listed in Table~\ref{Table-Fit-ZC-vs-R}. 
The results for $M_{\rm halo} = (10^{10} - 10^{11}) M_{\odot}$ 
fitted between $r = (1 - 40) h^{-1}$ kpc are: 
\begin{eqnarray} 
\label{eq-Fit-ZC-vs-R} 
\log \left( \frac{Z_C}{Z_{C, \odot}} \right) 
& = & - 0.209 + 2.23  (\log r) - 3.32 (\log r)^2 ~~ \textrm{NW},   \nonumber \\ 
& = & - 0.180 + 0.487 (\log r) - 1.54 (\log r)^2 ~~ \textrm{RVWa}  \nonumber. 
\end{eqnarray} 

We infer from Fig.~\ref{fig-rho-T-ZC-fC4-vs-R-MassBins}, third row 
that the CGM gas at galactocentric distances        
close to and beyond $R_{200}$, within $r \sim (30 - 300) h^{-1}$ kpc comoving, 
around galaxies of masses $M_{\rm halo} / M_{\odot} = 10^{9} - 10^{11}$, 
can give the best $Z_C$ observational diagnostic 
to distinguish between different galactic outflow models. 
Our metallicity dashed curves (enriched-particle only median, 
which is analogous to mass-weighted metallicity) are comparable to that of \citet{Hummels12}.

\subsubsection{CIV Fraction} 
\label{sec-res-Radial-CIV} 

We compute the fraction of triply ionized carbon, CIV, of the gas particles 
in post-processing using photoionization tables derived from CLOUDY 
\citep[last described by][]{Ferland98}. 
The relevant quantities used are redshift, gas density and temperature from simulation snapshot, 
and an ionizing background from the HM05 tables \citep{Haardt01}, 
as included in version 07.02.00 of CLOUDY. 

Fourth (bottom) row of 
Fig.~\ref{fig-rho-T-ZC-fC4-vs-R-MassBins} shows the radial profiles 
of the gas CIV fraction, $f_{\rm CIV}$. 
There are varying positive slopes at small-$r$ where $f_{\rm CIV}$ increases with radius. 
$f_{\rm CIV}$ attains a peak at an intermediate-$r$, 
which occurs at $(20 - 60) h^{-1}$ kpc depending on halo mass and wind model. 
Further out $f_{\rm CIV}$ decreases from its peak at larger-$r$, 
and have a negative slope up to $300 h^{-1}$ kpc. 

In the left two panels, runs CW and RVWa produce higher $f_{\rm CIV}$ than NW and RVWb; 
the differences are significant at $r \leq 7 h^{-1}$ kpc reaching $10^2 - 10^4$ times, 
very small at larger $r$ up to the peak of $f_{\rm CIV}$, 
and increases again to $10 - 70$ times at $r > 20 h^{-1}$ kpc. 
The small-$r$ positive-sloped regions in the right two panels show small differences: 
in the third RVWa has higher $f_{\rm CIV}$ than NW and RVWb, 
which are in turn higher than CW; 
whereas in the fourth (right) NW has higher $f_{\rm CIV}$ than CW, RVWa and RVWb. 
In the third panel $f_{\rm CIV}$ in runs CW and RVWa reaches a 
$\sim 50$ times higher peak at a larger-$r$ than runs NW and RVWb. 
The large-$r$ negative-sloped regions shows the same trends as the left 2 panels, 
runs CW and RVWa producing higher $f_{\rm CIV}$ than NW and RVWb. 

We perform polynomial fits to the $f_{\rm CIV}$ median radial profiles, 
in a similar way as in \S\ref{sec-res-Radial-ZC}, but with 2 first-order polynomials of the form: 
$\log f_{\rm CIV} = A + B (\log r)$, within 2 radial limits. 
The resulting fit coefficients are listed in Table~\ref{Table-Fit-fCIV-vs-R}. 
The polynomials for $M_{\rm halo} = (10^{10} - 10^{11}) M_{\odot}$, 
fitted between $r = (1 - 6) h^{-1}$ kpc are: 
\begin{eqnarray} 
\label{eq-Fit-fCIV-vs-R-Low} 
\log f_{\rm CIV} 
& = & - 6.58 + 5.18 (\log r) ~~ \textrm{NW},   \nonumber \\ 
& = & - 3.11 + 1.01 (\log r) ~~ \textrm{RVWa}; 
\end{eqnarray} 
and fitted within $r = (6 - 20) h^{-1}$ kpc are: 
\begin{eqnarray} 
\label{eq-Fit-fCIV-vs-R-High} 
\log f_{\rm CIV} 
& = & - 5.95 + 3.90 (\log r) ~~ \textrm{NW},   \nonumber \\ 
& = & - 5.43 + 3.67 (\log r) ~~ \textrm{RVWa}. 
\end{eqnarray} 

The overall behavior of $f_{\rm CIV}$ arises from the combined effects of 
density and temperature radial dependence of the gas 
(from first and second rows in Fig.~\ref{fig-rho-T-ZC-fC4-vs-R-MassBins}) 
in the ionization tables, with the major role played by the temperature. 

As a prediction for CIV fraction observations, 
we infer from Fig.~\ref{fig-rho-T-ZC-fC4-vs-R-MassBins}, bottom row that the 
inner gas at galactocentric distances $r < (4 - 5) h^{-1}$ kpc comoving, 
around galaxies of masses $M_{\rm halo} / M_{\odot} = 10^{10} - 10^{11}$, 
can most-effectively distinguish between strong-wind and no-wind cases.

\subsection{Metallicity-Density Relation} 
\label{sec-res-ZC-rho}

\begin{figure*} 
\centering 
\includegraphics[width = 1.0 \linewidth]{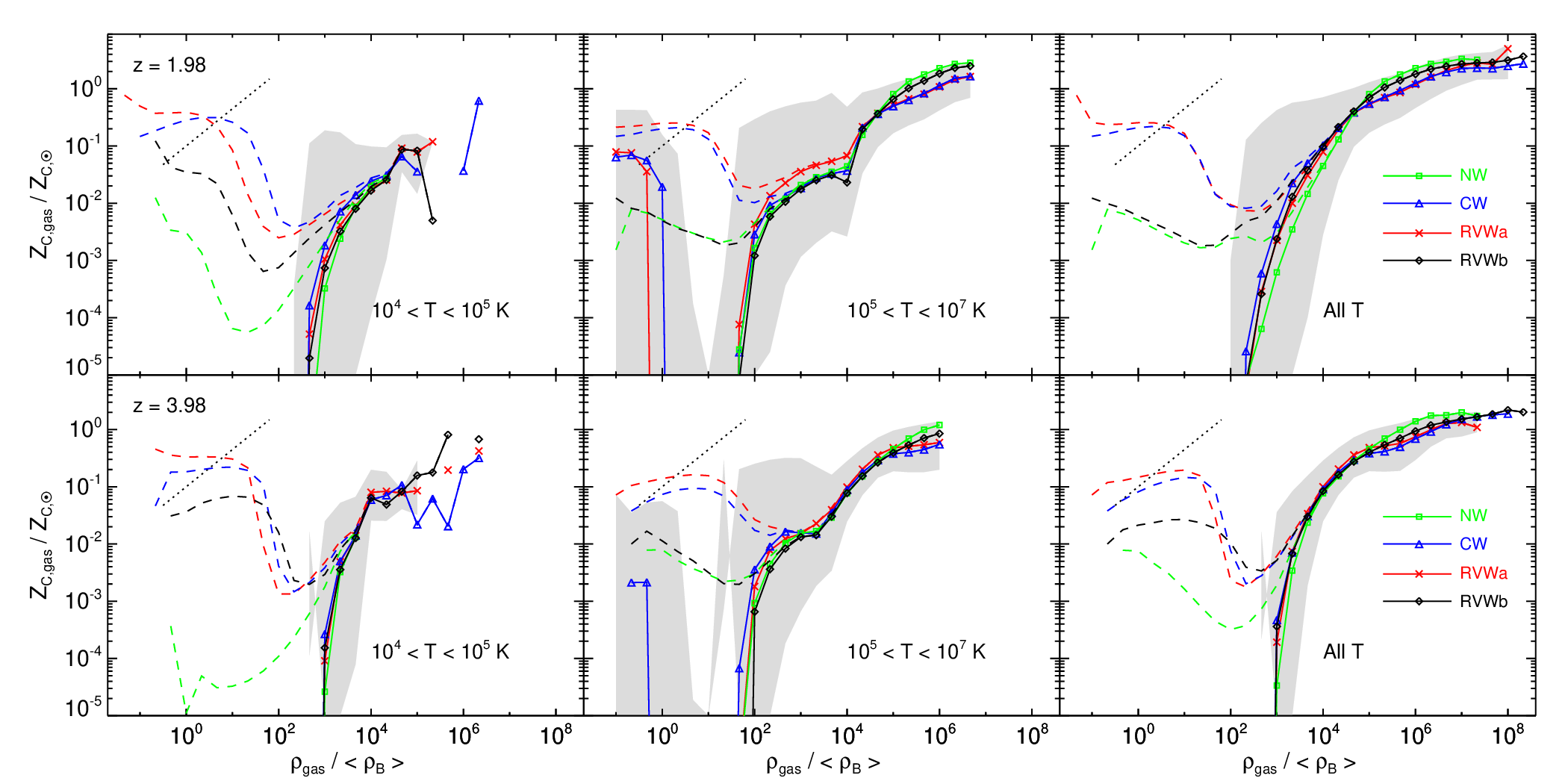} 
\caption{ 
Carbon metallicity versus density contrast of gas at 
$z = 1.98$ (top row, described in \S\ref{sec-res-ZC-rho}),   
and at an earlier epoch $z = 3.98$ (bottom row, discussed in \S\ref{sec-res-highz}),  
for the LB runs labeled by the color and plotting symbol. 
Gas in three temperature ranges are shown: 
$10^4 - 10^5$ K (left panels), $10^5 - 10^7$ K (middle), and all the gas (right). 
The plotted solid curves denote the median value in each density bin. 
The grey shaded area encloses the $70$th percentiles above and below the median in run RVWa 
(red curve), showing the typical scatter. 
These medians and percentiles are computed using all (both enriched and non-enriched) gas particles. 
The dashed curves show the median-$Z_C$ for enriched ($Z_C > 0$) particles only 
in density bins.    
The black dotted line in each panel shows 
the $\delta$-range and $[Z_C - \delta]$ slope    
obtained from observations by \citet{Schaye03}. 
} 
\label{fig-ZC-vs-Rho-TempBins} 
\end{figure*}

The carbon metallicity as a function of gas density contrast is plotted in 
Fig.~\ref{fig-ZC-vs-Rho-TempBins}, at $z = 1.98$ in the top row, 
for the LB runs labeled by the color and plotting symbol. 
The larger boxsize is selected again for increased statistics, 
since more number of halos and more massive halos form in it. 
Gas particles in three temperature bins are shown: 
$10^4 - 10^5$ K in the top-left panel, $10^5 - 10^7$ K in the top-middle, 
and all the gas in the top-right. 
The solid curves denote the median $Z_C$ in $\delta$-bins for each run. 
The grey shaded area enclose the $70$th percentiles above and below the median 
in run RVWa (red curve), showing the scatter at a given density. 
Following a format similar to Fig.~\ref{fig-rho-T-ZC-fC4-vs-R-MassBins}, 
the solid curve medians and percentiles are computed considering all 
(both enriched and non-enriched) gas particles, 
while the dashed curves represent median-$Z_C$ for the enriched particles only 
i.e. those having $Z_C > 0$. 
Note that if $\geq 50 \%$ of the particles are not enriched, 
then the solid median would be zero. 


The dashed median $Z_C$ in Fig.~\ref{fig-ZC-vs-Rho-TempBins}, top row, coincide with the 
solid median $Z_C$ at high densities, $\delta \geq 10^{3} - 10^{4}$, 
because most of the dense gas is star forming, generating metals and hence enriched. 
While at $\delta < 10^{3}$ the dashed median $Z_C$'s are several orders of magnitude 
higher than the solid ones (most clear in the top-left and top-right panels), 
implying that most or all of the low-density and underdense gas is not enriched. 

As an exception, the underdense gas having $0.1 \leq \delta \leq 2$ 
in the warm-hot $10^5 - 10^7$ K phase (top-middle panel) is
significantly enriched by the winds in runs CW and RVWa, so that 
all the particles (solid curves) have a median of $Z_C / Z_{C, \odot}
\sim 0.05 - 0.08$, just few times below the enriched-only dashed
medians.  Further trends discussed below analyzes the dashed medians
wherever the solid medians are below the plotting range.

Overall the metallicity-density relation shows a negative correlation at small-$\delta$ 
and positive correlation at large-$\delta$. 
Similar $[Z_C - \delta]$ relations were also found by \citet{Tornatore10} 
for the average metallicity of the warm-hot IGM at $z = 0$. 
We find that the threshold density for the slope turnover depends on 
temperature bin and wind model, lying between $\delta = 30 - 1000$. 
As a special feature, run CW has a slight increase in $Z_C$ from the smallest-$\delta$ to 
$\delta \sim 10$ in all the temperature bins, 
while run RVWa first decreases and then remains constant over the same range of densities. 

The wind models produce significantly different $Z_C$ in the low-density IGM at 
$\delta < 10^{2} - 10^{3}$, 
with runs CW and RVWa enriching up to few $100$ times higher than NW and RVWb. 
The metallicity induced by the four runs becomes similar at 
$\delta > 10^{4}$ in regions composed of SF gas. 
The warm phase of $10^4 - 10^5$ K (top-left) presents the largest differences 
between the wind models: 
CW enriches the $\delta \sim 30$ gas about $10$ times more than RVWa, 
which is $10$ times more than RVWb, which in turn is $50$ times more than NW. 

RVWa produces higher $Z_C$ than CW at $\delta < 5$ 
(top-left panel), but for larger-$\delta$ values CW enriches more. 
The same trend (RVWa enriching more than CW at small $\delta$) 
is visible in the other two panels (top-middle and top-right), 
but the resulting metallicity becomes almost comparable in RVWa and CW at larger densities. 

Applying the pixel optical depth technique to quasar spectra, 
\citet{Schaye03} found a positive gradient of observed CIV metallicity with density contrast, 
measuring a median $[C/H] \propto 0.65 \log \delta$, 
in the range $\log \delta = [-0.5, 1.8]$ and $z = 1.8 - 4.1$. 
Comparing with our Fig.~\ref{fig-ZC-vs-Rho-TempBins}, dashed curves in the top row, 
some of the wind runs show a weak similar trend in this $\delta$ range. 
In all the top panels of run CW and top-middle panel of RVWa, $Z_C$ show a weak positive correlation 
with $\delta$, but with a slope $\sim 4$ times smaller than observed. 
On the other hand, runs NW and RVWb, top-left and top-right panels of RVWa, 
along with the all-particle solid medians of runs CW and RVWa in the top-middle panel, 
show a negative correlation of $Z_C$ with $\delta$, or almost constant $Z_C$. 
At the same time, 
this negative correlation is consistent with the results of \citet{Barai11}. 
Studying IGM enrichment from anisotropic AGN outflows, 
they found that at $z \geq 2$ the underdense regions ($\delta < 1$) are enriched 
to higher metallicities, and the resulting $[O/H]$ decreases with increasing IGM density, 
a trend more prominent with increasing anisotropy of the outflows. 

We perform polynomial fits to the median $Z_C$ (solid curves) 
versus $\delta$ data of runs NW and RVWa, 
in a similar way as in \S\ref{sec-res-Radial-ZC} and \S\ref{sec-res-Radial-CIV}, 
fitting with 2 first-order polynomials of the form: 
$\log (Z_C / Z_{C, \odot}) = A + B (\log \delta)$, within 2 density limits. 
The resulting fit coefficients are listed in Table~\ref{Table-Fit-ZC-vs-Rho}, 
for the 3 temperature ranges plotted in Fig.~\ref{fig-ZC-vs-Rho-TempBins}. 
The polynomials in the warm-hot $(10^5 - 10^7)$ K gas, 
fitted within $10^{2} \leq \delta < 10^{4}$ are: 
\begin{eqnarray} 
\label{eq-Fit-ZC-vs-Rho-Low} 
\log \left( \frac{Z_C}{Z_{C, \odot}} \right) 
& = & - 3.81 + 0.656 (\log \delta) ~~ \textrm{NW},   \nonumber \\ 
& = & - 3.22 + 0.546 (\log \delta) ~~ \textrm{RVWa}; 
\end{eqnarray} 
and fitted between $10^{4} \leq \delta < 10^{7}$ are: 
\begin{eqnarray} 
\label{eq-Fit-ZC-vs-Rho-High} 
\log \left( \frac{Z_C}{Z_{C, \odot}} \right) 
& = & - 3.55 + 0.643 (\log \delta) ~~ \textrm{NW},   \nonumber \\ 
& = & - 2.70 + 0.456 (\log \delta) ~~ \textrm{RVWa}. 
\end{eqnarray}

\subsection{Redshift Evolution: Metallicity at $z \sim 4$} 
\label{sec-res-highz} 

We investigate the redshift evolution of two metallicity correlations: 
$Z_C$ versus $r$, and $Z_C$ versus $\delta$, of the LB runs. 

\begin{figure*} 
\centering 
\includegraphics[width = 0.95 \linewidth]{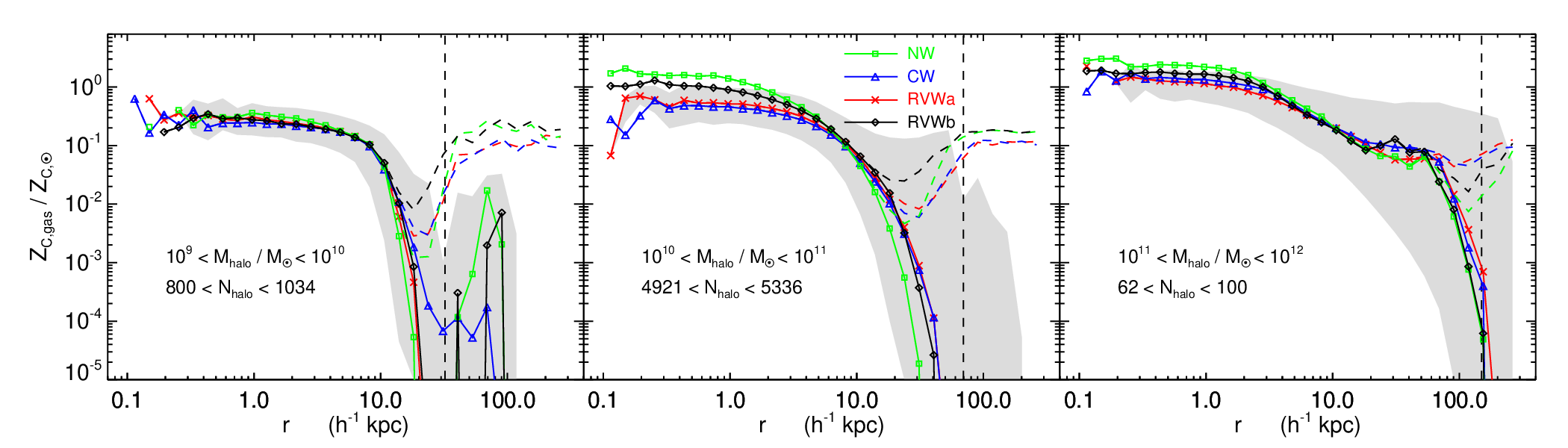} 
\caption{ 
Gas carbon metallicity radial profile of galaxies at $z = 3.98$,   
an earlier epoch than in Fig.~\ref{fig-rho-T-ZC-fC4-vs-R-MassBins}, 
third row, in a similar format. 
It shows the ratio of carbon mass fraction in gas to that of the Sun, 
as a function of comoving radius, for 3 halo mass ranges. 
The dashed curves represent median $Z_C$ values for the enriched particles only 
in each radial bin. See \S\ref{sec-res-highz}. 
} 
\label{fig-ZC-vs-R-MassBins-005} 
\end{figure*}

The gas carbon abundance radial profiles of galaxies at $z = 3.98$ 
are plotted in Fig.~\ref{fig-ZC-vs-R-MassBins-005}, 
which is an earlier epoch than Fig.~\ref{fig-rho-T-ZC-fC4-vs-R-MassBins}. 
The 3 panels in Fig.~\ref{fig-ZC-vs-R-MassBins-005} denote 3 total halo mass ranges, 
$M_{\rm halo} / M_{\odot} = 10^{9} - 10^{10}$ (left panel) with 
number of halos in the 4 plotted runs between $N_{\rm halo} = 800 - 1034$, 
$10^{10} - 10^{11}$ (middle) having $N_{\rm halo} = 4921 - 5336$, 
and $10^{11} - 10^{12}$ (right) where $N_{\rm halo} = 62 - 100$. 
Because of the earlier time, here no halo has grown to the highest bin mass 
of the previous radial profile plots at $z = 1.98$ (\S\ref{sec-res-Radial-Profile}). 



We see that several of the median $Z_C$ versus $r$ correlations 
are similar between $z = 3.98$ and $z = 1.98$, 
only the absolute metallicity values are lower at the earlier epoch. 
All the runs at $z = 3.98$ have $Z_C$ decreasing with $r$ at $r \leq (0.5 - 2) R_{200}$, 
with $r$-dependent negative slopes. 
The trend of $Z_C$ profiles rising again at larger $r$ values 
(because of in-situ metal generation in surrounding structures and 
spreading of metals by wind from the central SF regions) is weaker at $z = 3.98$; 
it is only visible for the 
dashed curves in all the panels and few solid curves in the left panel. 

The differences between the wind models are smaller at $z = 3.98$ than $z = 1.98$, 
however there are signatures of suppression of central SF by the impact of winds, 
which transport metals out and accumulate them in the CGM. 
Analyzing the solid curves in the middle and right panels, 
runs CW and RVWa have few times lower $Z_C$ than NW and RVWb in the inner $r < (5 - 7) h^{-1}$ kpc. 
The trend reverses in the outer $r > (5 - 7) h^{-1}$ kpc; 
CW, RVWa and RVWb have few times higher metallicity than NW. 
The enriched-only dashed curves show a feature 
different from their solid counterparts in the left and middle panels, 
for large $r$ values, where they detach from the solid curves and increases with $r$; 
$Z_C$ in runs NW and RVWb are $\sim 2 - 3$ times higher than CW and RVWa.

The bottom row of Fig.~\ref{fig-ZC-vs-Rho-TempBins} shows the gas carbon metallicity 
versus density contrast at $z = 3.98$, for 3 temperature ranges, following the top row. 
Analogous to $[Z_C - r]$, several of the median $Z_C$ versus $\delta$ correlations 
at $z = 3.98$ are similar to those at $z = 1.98$, 
with the absolute metallicity values being lower at earlier times. 
The differences between the wind models are smaller at $z = 3.98$ than $z = 1.98$. 

The underdense gas with $0.1 \leq \delta \leq 0.8$ in the warm-hot $10^5 - 10^7$ K phase 
(bottom-middle panel) is already enriched by the winds in run CW, 
making the all-particle (solid curve) median $Z_C / Z_{C, \odot} \sim 0.002$ at $z = 3.98$, 
which is $40$ times lower than at $z = 1.98$. 
However, the enrichment at similar $\delta$ values caused by the winds of 
model RVWa at $z = 3.98$ is substantially lower than that of CW by $1000$ times or more. 

Further trends at $z = 3.98$ discussed below analyze the dashed medians 
wherever the solid medians are below the plotting range. 
Starting from the lowest-$\delta$, 
the metallicity-density in all the outflow runs (except RVWb in the bottom-middle panel) 
show a shallow positive correlation at $\delta \leq 30$. 
At $z = 1.98$, such a feature is only present in run CW. 
This positive-sloped $[Z_C - \delta]$ relation that we obtain at $z = 3.98$ in the wind models 
is consistent with observations by \citet{Schaye03} who found a positive gradient of 
CIV metallicity at similar overdensities.

\section{Summary and Conclusion} 
\label{sec-conclusion} 

We explore new models of galactic winds performing hydrodynamic simulations 
using the TreePM-SPH code {\sc GADGET-3}, 
and analyze their impact on the properties of the CGM at $z = 2 - 4$. 
Our outflow implementation imparts kinetic feedback, in the
energy-driven formalism, where the wind velocity has a positive
correlation with galactocentric radius $v_w(r)$ as seen in
observations by \citet{Steidel10}.  We further investigate a halo mass
dependent parametrization of the radially-varying wind, following 
observations by \citet{Martin05}.  The simulations include additional 
subgrid physics: metal-dependent radiative cooling and heating in the
presence of photoionizing background radiation; star formation;
stellar evolution and self-consistent chemical enrichment using a fixed stellar IMF. 

The implementation of the new wind models in the code 
involves finding the distance of gas particles
from their host galaxy center.  We identify galaxies by running a
FOF group finder on-the-fly within a simulation at intervals of
$1.001$ times the scale factor, 
and find stellar groups of at least $32$ particles, 
by linking over stars as the primary particle type,
using a linking length 3 times smaller than that for obtaining DM
halos.  The location of the member gas particle with maximum density
is considered as the galaxy center.  Multiphase gas particles
undergoing SF are stochastically selected and kicked into wind by
giving their speed a one-time $v_w$ boost in a direction 
perpendicular to the galaxy disk. 
Wind particles are also temporarily decoupled from hydrodynamical interactions. 

We simulate two different cosmological volumes (in order to increase the statistics)
a smaller box of $(5 h^{-1}$ Mpc$)^3$ comoving with $2 \times 128^3$ DM and gas particles, 
and a larger box of $(25 h^{-1}$ Mpc$)^3$ with $2 \times 320^3$ particles; 
using the flat $\Lambda$CDM concordance model. 
For each volume, we perform 4 runs investigating different galactic wind models: \\ 
NW: no wind; \\ 
CW: energy-driven wind with constant $v_w = 400$ km/s; \\ 
RVWa: radially varying wind with fixed parameters; \\ 
RVWb: RVW with parameters dependent on halo mass. 

The main analyses of our simulations reflect the following processes:
the outflows expel gas away from the star-forming galaxies and
suppress the SF; the gas is also metal-enriched; thus winds carry
metals out and accumulate them in the CGM and IGM, enriching these
lower density regions with metals. Our results are summarized below. 

\begin{itemize} 

\item[-] {\it Outflow speed.} 

The outflow gas velocity magnitude as a function of galactocentric distance 
of multiple wind-phase gas particles, obtained in our simulations, 
follows the given input subgrid wind speed, constant or    
varying with radius, in agreement with observations. 

\item[-] {\it Star formation rate.} 

Galactic wind feedback quenches SF at $z < 8$. 
In the $(25 h^{-1}$ Mpc$)^3$ runs at $z \sim 2$, the global SFRD is 
$4$ times smaller in RVWa and $2$ times smaller in CW than NW. 
At $z \leq 5$ RVWa causes a greater suppression than CW, 
and produces $2 - 4$ times lower SFRD than CW at $z = 2$. 
The SFRD at $z \sim 4.5 - 10$ is systematically (up to $2 - 10$ times) 
higher in the simulations, compared to observations. 
At later epochs $z \sim 2 - 4.5$, most of the observations \citep[e.g.,][]{Cucciati12} 
lie within the ranges of SFRD produced by the different wind models. 

The sSFR versus galaxy stellar mass at $z \sim 2$ 
present reasonably good agreement with observations, the models reproducing 
the observed main sequence \citep[e.g.,][]{Daddi07} for star forming galaxies.

\item[-] {\it Gas and stellar mass functions and mass fractions.} 

  The gas mass function
  of galaxies in the $(25 h^{-1}$ Mpc$)^3$ runs at $z = 2.23$ has the
  same slope between $M_{\rm gas} \sim (10^{9} - 10^{11}) M_{\odot}$,
  but in CW and RVWa is shifted leftward with respect to NW and RVWb by $(2 - 3)
  M_{\odot}$; because winds expel gas, causing a smaller number 
  of objects with high gas masses. 
  Runs CW and RVWa produce a steeper stellar mass function than NW and RVWb. 
  There is a small excess of galaxies in RVWa compared to CW between 
  $M_{\rm stellar} = (2 \times 10^{8} - 10^{9}) M_{\odot}$. 
  Our model RVWa provide a reasonably good match to the observational data 
  of the stellar mass function at $2 \leq z < 3$ \citep[e.g.,][]{Marchesini09}, 
  over $M_{\rm stellar} = (8 \times 10^{9} - 10^{11}) M_{\odot}$. 
  
In run NW the gas mass fraction decreases monotonically with halo mass; 
outflows (runs CW, RVWa, RVWb) flatten the $M_{\rm gas} / M_{\rm halo}$ trend 
making it oscillatory. 
The stellar mass fraction in galaxies is the largest in run NW, 
and increases with $M_{\rm halo}$, 
such that NW and RVWb have the steepest increase, followed by CW, and RVWa is the flattest.

\item[-] {\it Thermal state of the gas.} 

The $[T - \rho]$ phase diagram shows that the outflow, soon after 
leaving the dense SF phase, goes through a cold ($< 10^4$ K) phase,
due to the hydrodynamic decoupling.  The thermal properties of the IGM
look qualitatively similar in the 4 runs.  Model RVWa has a higher
fraction of underdense to slightly overdense ($\delta \sim 0.1 - 10$),
warm-hot ($T \sim 10^4 - 10^6$ K) gas, than CW; both of them are
significantly higher than NW and RVWb.

\item[-] {\it Gas kinematics.} 

Projection of gas kinematics  around the center of a galaxy of 
total halo mass $M_{\rm halo} \sim 2 \times 10^{11} M_{\odot}$ at $z = 2.12$, 
shows that for NW and RVWb the gas outflows 
along the direction of least resistance of a low-density void. 
Run CW has most of the outflow inside $r \leq 50$ kpc forming a metal-enriched gas sphere, 
$v_w = 400$ km/s not being enough to drive the wind to larger distances. 
The $v_w(r)$ in run RVWa produces an extended diffuse enriched gas outflow 
propagating perpendicular to the galaxy disk (as expected from the model input), 
escaping to $r > 100$ kpc. 
The formation of a galaxy disk is visible in runs NW, RVWa and RVWb, 
where inside the $(200 h^{-1}$ kpc$)^3$ projection volume 
about half of the gas is inflowing and the other half outflowing. 
Whereas the CW galaxy looks irregularly shaped, 
and more gas $(64 \%)$ is undergoing infall than outflow $(36 \%)$. 
The galaxy disk in run RVWa is the largest in size and further investigations are needed 
to check if $v_w(r)$ can produce realistic disk galaxies.

\item[-] {\it Radial profiles.} 

The radial profiles of gas properties around galaxy centers 
at $z = 1.98$ show most prominent differences between the models 
for lower halo masses ($M_{\rm halo} / M_{\odot} = 10^{9} - 10^{10}$ and $10^{10} - 10^{11}$), 
and almost uniform results at higher masses. 
The density and carbon metallicity profiles often form a local peak 
at $r > R_{200}$ of galaxies, 
because of the presence of smaller halos and surrounding substructures where 
more metals are produced by in-situ SF, and the spreading of metals by wind. 

\begin{itemize} 

\item[$\bullet$] {\it Gas density profile.} 
Runs CW and RVWa have a lower density in the inner $r \leq 10 h^{-1}$ kpc, 
by $10 - 30$ times, than the NW and RVWb cases. 
While in the outer $r > R_{200}$, CW, RVWa and RVWb show a higher density than NW. 

\item[$\bullet$] {\it Temperature profile.} 
The wind runs CW, RVWa and RVWb present a bimodal temperature distribution 
at $r \leq 10 h^{-1}$ kpc, composed of hot multiphase star forming gas and cold winds. 
Between $(10 - 300) h^{-1}$ kpc, the NW model cause the highest-$T$ 
($\sim 10^{5} - 10^{6}$ K), followed by RVWb, then RVWa and CW. 

\item[$\bullet$] {\it Carbon metallicity profile.} 
Runs CW and RVWa have a lower $Z_C$ than NW and RVWb 
in the inner $r < (5 - 7) h^{-1}$ kpc, while in the outer parts 
runs CW, RVWa and RVWb produces higher $Z_C$ by $20 - 30$ times than NW. 
Metallicity in RVWb becomes larger than in NW 
at $r > 5 h^{-1}$ kpc and up to $10$ times higher at $r \geq R_{200}$. 
We perform second order polynomial fits to $Z_C$ versus $r$, 
whose coefficients are listed in Table~\ref{Table-Fit-ZC-vs-R}. 

The carbon enriched ($Z_C > 0$) gas fraction is $1$ in the inner regions of galaxies; 
it reduces to $(0.5 - 0.8)$ at $R_{200}$ and to $(0.45 - 0.7)$ at $300 ~ h^{-1}$ kpc. 
Runs CW and RVWa enrich a higher (up to $1.4$ times) fraction of gas than NW and RVWb. 

\item[$\bullet$] {\it CIV fraction around galaxies.} 
CIV fraction profiles have varying positive and negative slopes at different $r$, 
attaining a peak between $(20 - 60) h^{-1}$ kpc. 
Runs CW and RVWa produces higher $f_{\rm CIV}$ than NW and RVWb. 
We fit the $f_{\rm CIV}$ radial profiles with 2 first-order polynomials, 
and list the coefficients in Table~\ref{Table-Fit-fCIV-vs-R}. 
The results for $M_{\rm halo} = (10^{10} - 10^{11}) M_{\odot}$, 
fitted between $r = (1 - 6) h^{-1}$ kpc show the largest difference: 
$f_{\rm CIV} \propto r^{6}$ for NW, and 
$f_{\rm CIV} \propto r^{1}$ for RVWa.

\item[$\bullet$] {\it Observational Predictions.} 
Inferred from our simulations, we predict that $Z_C$ observations of the CGM gas 
at galactocentric distances in the range $r \sim (30 - 300) h^{-1}$ kpc comoving, 
around galaxies of $M_{\rm halo} = (10^{9} - 10^{11}) M_{\odot}$, 
can best distinguish between different galactic outflow scenarios. 
And CIV fraction observations of the inner gas in the range $r < (4 - 5) h^{-1}$ kpc comoving, 
around galaxies of $M_{\rm halo} = (10^{10} - 10^{11}) M_{\odot}$, 
can most-effectively distinguish between strong-wind and no-wind cases. 

\end{itemize}

\item[-] {\it Metallicity as a function of density.} 

The underdense $\delta = 0.1 - 2$ warm-hot $10^5 - 10^7$ K phase 
is significantly enriched by the winds in runs CW and RVWa, 
so that the all-particles median is $Z_C / Z_{C, \odot} \sim 0.02 - 0.08$. 
$Z_C$ versus $\delta$ at $z = 1.98$ shows a negative correlation at small-$\delta$, 
and positive correlation at large-$\delta$. 
The only exception is run CW where $Z_C$ has a small increase between $\delta \sim 0.1 - 10$, 
while $Z_C$ in RVWa first decreases and then remains flat. 
The low-density IGM with $\delta < 10^{2} - 10^{3}$ is significantly enriched 
in runs CW and RVWa, up to few $100$ times more than NW and RVWb. 
In the warm $10^4 - 10^5$ K phase, CW enriches the $\delta \sim 30$ IGM 
$10$ times more than RVWa, which enriches $10$ times more than RVWb, 
which in turn enriches $50$ times more than NW. 

We perform 2 first-order polynomial fits to the $[Z_C - \delta]$ correlation, 
whose coefficients are listed in Table~\ref{Table-Fit-ZC-vs-Rho}. 
As an example, for the warm-hot $(10^5 - 10^7)$ K gas, 
the fits within $10^{2} \leq \delta < 10^{4}$ are: 
$Z_C \propto \delta^{0.656}$ for NW, and 
$Z_C \propto \delta^{0.546}$ for RVWa.

\item[-] {\it Redshift evolution.} 

Several of the $[Z_C - r]$ and $[Z_C - \delta]$ correlations 
at $z = 3.98$ are similar to those at $z = 1.98$, 
with lower metallicity values and 
smaller differences between the wind models at the earlier epoch. 
The trend of $Z_C$ radial profiles rising again at large-$r$ is almost absent at $z = 3.98$. 
For $M_{\rm halo} / M_{\odot} = 10^{9} - 10^{10}$ and $10^{10} - 10^{11}$, 
the enriched-only median $Z_C$ 
in runs NW and RVWb at large-$r$ are $2 - 3$ times higher than in CW and RVWa. 

The $z = 3.98$ underdense $\delta = 0.1 - 0.8$ warm-hot $10^5 - 10^7$ K gas 
is enriched by the winds in run CW (but not in RVWa) making 
the all-particle median $Z_C / Z_{C, \odot} \sim 0.002$, which is $40$ 
times lower than at $z = 1.98$.  The $[Z_C - \delta]$ relation at $z = 3.98$ in 
all the wind runs (except RVWb in the warm-hot phase) shows a shallow 
positive correlation over $\delta \sim 0.1 - 30$, consistent with 
observations \citep{Schaye03}. 

\end{itemize} 

In summary, we have found that the wind model with the radially varying velocity
dependent on halo mass (RVWb) is the least effective in modifying IGM 
related properties, with results similar to the no-wind
(NW) case, except that it substantially enriches the low-density CGM.

The impact of the model RVWa, for which the velocity is
increasing as a function of galactocentric distance, is instead 
similar to the energy-driven constant-velocity implementation CW. 
However, it shows interesting differences that deserve to be further investigated. 
RVWa causes a greater suppression of SFR than CW at $z \leq 5$, 
this could have implications for the galaxy downsizing scenario. 
RVWa also produces galactic disks larger than all the other wind models, 
and one can study if the radially varying outflow formalism can generate more
realistic disk galaxies. 
Run RVWa has a higher gas fraction than run CW in
the low-density ($\delta \sim 0.1 - 10$) warm-hot ($10^4 - 10^6$ K)
phase of the IGM, which could shed light on the missing baryon
problem. 

We see different trends of $Z_C$ versus $\delta$ between the CW and
RVWa models at $\delta \leq 10$, and CW outflows generally produce a
higher and earlier enrichment of some IGM phases than RVWa.  To 
explore such IGM metal-enrichment differences, future progress in this 
field should include computing more observable statistics from the 
simulations, e.g., Lyman-$\alpha$ flux, simulated quasar spectra, and 
compare them with observations of CGM and IGM at different impact 
parameters from galaxies.

\section*{Acknowledgments} 

We are grateful to Volker Springel for allowing us to use the GADGET-3 code. 
Calculations for this paper were partly performed on the COSMOS
Consortium supercomputer within the Dirac Facility jointly funded by
STFC, the Large Facilities Capital Fund of BIS and the University of
Cambridge, as well as the Darwin Supercomputer of the University of
Cambridge High Performance Computing Service
(http://www.hpc.cam.ac.uk/), provided by Dell Inc. using Strategic
Research Infrastructure Funding from the Higher Education Funding
Council for England. Simulations were also run at the CINECA Super
Computer Center (CPU time assigned through an INAF-CINECA grant).  We
thank Olga Cucciati for sending us observational data for SFRD, and 
Martin Haehnelt, Giuseppe Murante 
for useful discussions. This work is supported by
PRIN-MIUR, PRIN-INAF 2009, INFN/PD51 grant. 
PB and MV are supported by the ERC Starting Grant ``cosmoIGM''. 
ET acknowledges the Australian Research Council Centre of Excellence 
for All-sky Astrophysics (CAASTRO), funded by grant CE110001020. 
MK acknowledges a fellowship from the European Commission's Framework Programme 7, 
through the Marie Curie Initial Training Network CosmoComp (PITN-GA-2009-238356). 

%

\appendix 

\section{} 


%


\begin{table*} 
\begin{minipage}{12cm} 
\caption{ 
Polynomial Fit of Carbon Metallicity Radial Profile. 
Coefficients from fitting median $Z_C$ versus $r$ data with a 
second order polynomial of the form, $\log (Z_C / Z_{C, \odot}) = A + B (\log r) + C (\log r)^2$. 
Fits are calculated for 2 runs from Fig. \ref{fig-rho-T-ZC-fC4-vs-R-MassBins}, 
third row, at $z = 1.98$, 
for the 4 halo mass ranges (4 panels in the figure) within radial ranges indicated here. 
Columns 1, 2: Minimum and maximum halo mass indicating the range of the median $Z_C$ result. 
Columns 3, 4: Minimum and maximum radius from galaxy center within which the fit is done. 
Column 5: Name of simulation run. 
Columns 6, 7, 8: Fitting coefficients $A$, $B$, $C$. 
} 

\label{Table-Fit-ZC-vs-R} 

\begin{tabular}{cccccccc} 
\hline 
$M_{\rm halo, min}$ & 
$M_{\rm halo, max}$ & 
$r_{\rm min}$ & 
$r_{\rm max}$ & 
Run & 
\multicolumn{3}{c}{Fit Coefficients} \\ 

[$M_{\odot}$] & 
[$M_{\odot}$] & 
[$h^{-1}$ kpc] & 
[$h^{-1}$ kpc] & 
Name & 
$A$ & 
$B$ & 
$C$ \\ 
\hline



$10^{9}$ & $10^{10}$ & $1$ & $30$ & NW   & $0.578$  & $-0.685$ & $-1.60$ \\ 
$10^{9}$ & $10^{10}$ & $1$ & $30$ & RVWa & $-0.519$ & $1.08$   & $-1.85$ \\ \\ 

$10^{10}$ & $10^{11}$ & $1$ & $40$ & NW   & $-0.209$ & $2.23$  & $-3.32$ \\  
$10^{10}$ & $10^{11}$ & $1$ & $40$ & RVWa & $-0.180$ & $0.487$ & $-1.54$ \\ \\ 

$10^{11}$ & $10^{12}$ & $1$ & $300$ & NW   & $0.227$  & $0.583$ & $-1.03$ \\  
$10^{11}$ & $10^{12}$ & $1$ & $300$ & RVWa & $-0.265$ & $1.02$  & $-1.03$ \\ \\ 

$10^{12}$ & $10^{13}$ & $1$ & $300$ & NW   & $0.373$ & $-0.0354$ & $-0.384$ \\ 
$10^{12}$ & $10^{13}$ & $1$ & $300$ & RVWa & $0.498$ & $-0.241$  & $-0.255$ \\ 


\hline
\end{tabular} 

\end{minipage}
\end{table*} 


%


\begin{table*} 
\begin{minipage}{15cm} 
\caption{ 
Polynomial Fit of CIV Fraction Radial Profile. 
Coefficients from fitting median $f_{\rm CIV}$ versus $r$ data with two 
first-order polynomials of the form: 
$\log f_{\rm CIV} = A_1 + B_1 (\log r)$ within $[r_{1, \rm min}, r_{1, \rm max}]$, and 
$\log f_{\rm CIV} = A_2 + B_2 (\log r)$ within $[r_{2, \rm min}, r_{2, \rm max}]$. 
Fits are calculated for 2 runs from Fig. \ref{fig-rho-T-ZC-fC4-vs-R-MassBins}, 
fourth row, at $z = 1.98$, 
for the 4 halo mass ranges (4 panels in the figure). 
Columns 1, 2: Minimum and maximum halo mass indicating the range of the median $f_{\rm CIV}$ result. 
Column 3: Name of simulation run. 
Columns 4, 5: Minimum and maximum radius from galaxy center within which first fit is done. 
Columns 6, 7: Fitting coefficients $A_1$, $B_1$. 
Columns 8, 9: Minimum and maximum radius from galaxy center within which second fit is done. 
Columns 10, 11: Fitting coefficients $A_2$, $B_2$. 
} 

\label{Table-Fit-fCIV-vs-R} 

\begin{tabular}{ccccccccccc} 
\hline 
$M_{\rm halo, min}$ & 
$M_{\rm halo, max}$ & 
Run & 
$r_{1, \rm min}$ & 
$r_{1, \rm max}$ & 
\multicolumn{2}{c}{Fit Coeffs-1} & 
$r_{2, \rm min}$ & 
$r_{2, \rm max}$ & 
\multicolumn{2}{c}{Fit Coeffs-2} \\ 

[$M_{\odot}$] & 
[$M_{\odot}$] & 
Name & 
[$h^{-1}$ kpc] & 
[$h^{-1}$ kpc] & 
$A_1$ & 
$B_1$ & 
[$h^{-1}$ kpc] & 
[$h^{-1}$ kpc] & 
$A_2$ & 
$B_2$ \\ 
\hline 



$10^{9}$ & $10^{10}$ & NW   & $1$ & $6$ & $-3.92$ & $2.18$  & $6$ & $20$ & $-5.27$ & $3.69$ \\ 
$10^{9}$ & $10^{10}$ & RVWa & $1$ & $6$ & $-2.68$ & $0.613$ & $6$ & $20$ & $-4.54$ & $3.13$ \\ \\ 

$10^{10}$ & $10^{11}$ & NW   & $1$ & $6$ & $-6.58$ & $5.18$ & $6$ & $20$ & $-5.95$ & $3.90$ \\ 
$10^{10}$ & $10^{11}$ & RVWa & $1$ & $6$ & $-3.11$ & $1.01$ & $6$ & $20$ & $-5.43$ & $3.67$ \\ \\ 

$10^{11}$ & $10^{12}$ & NW   & $2$ & $7$ & $-12.19$ & $9.52$ & $7$ & $20$ & $-4.87$ & $1.96$ \\ 
$10^{11}$ & $10^{12}$ & RVWa & $2$ & $7$ & $-10.63$ & $8.05$ & $7$ & $20$ & $-5.94$ & $2.90$ \\ \\ 

$10^{12}$ & $10^{13}$ & NW   & $4$ & $10$ & $-16.50$ & $13.29$ & $10$ & $30$ & $-3.78$ & $0.624$ \\ 
$10^{12}$ & $10^{13}$ & RVWa & $4$ & $10$ & $-14.13$ & $9.66$  & $10$ & $30$ & $-7.17$ & $3.10$ \\ 


\hline 
\end{tabular} 

\end{minipage} 
\end{table*} 


%


\begin{table*} 
\begin{minipage}{13cm} 
\caption{ 
Polynomial Fit of Carbon Metallicity-Density Contrast Relation. 
Coefficients from fitting median $Z_C$ versus 
$\delta = \rho_{\rm gas} / \langle \rho_B \rangle$ data with two first-order polynomials of the form: 
$\log (Z_C / Z_{C, \odot}) = A_1 + B_1 (\log \delta)$ within $[\delta_{1, \rm min}, \delta_{1, \rm max}]$, and 
$\log (Z_C / Z_{C, \odot}) = A_2 + B_2 (\log \delta)$ within $[\delta_{2, \rm min}, \delta_{2, \rm max}]$. 
Fits are calculated for 2 runs from Fig. \ref{fig-ZC-vs-Rho-TempBins} at $z = 1.98$, 
for the 3 given temperature ranges (3 panels in the figure). 
Columns 1, 2: Minimum and maximum temperature indicating the range of the median $Z_C$ result. 
Column 3: Name of simulation run. 
Columns 4, 5: Minimum and maximum density contrast within which first fit is done. 
Columns 6, 7: Fitting coefficients $A_1$, $B_1$. 
Columns 8, 9: Minimum and maximum density contrast within which second fit is done. 
Columns 10, 11: Fitting coefficients $A_2$, $B_2$. 
} 

\label{Table-Fit-ZC-vs-Rho} 

\begin{tabular}{ccccccccccc} 
\hline 
$T_{\rm min}$ & 
$T_{\rm max}$ & 
Run & 
$\delta_{1, \rm min}$ & 
$\delta_{1, \rm max}$ & 
\multicolumn{2}{c}{Fit Coeffs-1} & 
$\delta_{2, \rm min}$ & 
$\delta_{2, \rm max}$ & 
\multicolumn{2}{c}{Fit Coeffs-2} \\ 

[$^{\circ}$ K] & 
[$^{\circ}$ K] & 
Name & & & 
$A_1$ & 
$B_1$ & & & 
$A_2$ & 
$B_2$ \\ 
\hline 



$10^{4}$ & $10^{5}$ & NW   & $10^{3}$ & $10^{5}$ & $-7.54$ & $1.43$  & - & - & - & - \\ 
$10^{4}$ & $10^{5}$ & RVWa & $10^{3}$ & $10^{5}$ & $-5.59$ & $0.937$ & - & - & - & - \\ \\ 

$10^{5}$ & $10^{7}$ & NW   & $10^{2}$ & $10^{4}$ & $-3.81$ & $0.656$ & $10^{4}$ & $10^{7}$ & $-3.55$ & $0.643$ \\ 
$10^{5}$ & $10^{7}$ & RVWa & $10^{2}$ & $10^{4}$ & $-3.22$ & $0.546$ & $10^{4}$ & $10^{7}$ & $-2.70$ & $0.456$ \\ \\ 

$10^{2}$ & $10^{8}$ & NW   & $10^{3}$ & $10^{5}$ & $-7.62$ & $1.54$ & $10^{5}$ & $10^{8}$ & $-1.20$ & $0.247$ \\ 
$10^{2}$ & $10^{8}$ & RVWa & $10^{3}$ & $10^{5}$ & $-5.98$ & $1.19$ & $10^{5}$ & $10^{8}$ & $-1.74$ & $0.300$ \\ 


\hline 
\end{tabular} 

\end{minipage} 
\end{table*} 



\end{document}